\documentclass[sort & compress, review, 12pt]{elsarticle}

\usepackage{lineno,hyperref, appendix, booktabs}
\usepackage{amsmath, bm, amssymb}
\usepackage{longtable}
\usepackage{subfigure, float, overpic,graphicx}
\usepackage[a4paper, total={7in, 10in}]{geometry}
\DeclareUnicodeCharacter{03B2}{\ensuremath{\beta}}

\modulolinenumbers[10]

\journal{Journal of \LaTeX\ Templates}

\AtBeginDocument{%
  \def\thefnote{\myfnsymbol{fnote}}}
\makeatletter
\def\myfnsymbol#1{\expandafter\@myfnsymbol\csname c@#1\endcsname}
\def\@myfnsymbol#1{\ifcase #1\or $\ast$\or $\#\#$\else \@ctrerr\fi}
\def\fntext[#1]#2{\g@addto@macro\@fnotes{%
   \refstepcounter{fnote}\elsLabel{#1}%
   \def\thefootnote{\thefnote}
   \global\setcounter{footnote}{\c@fnote}%
   \footnotetext{#2}}}
\makeatother

\begin{document}

\begin{frontmatter}

\title{Investigating amorphization as a deformation mechanism using a novel phase field 
model at the mesoscale}
\author[mymainaddress]{Yuntong Huang}

\author[secondaddress]{Shuyang Dai \fnref{fn1}}
\ead{shuyang_dai@whu.edu.cn}

\author[mymainaddress]{Chuqi Chen}

\author[mymainaddress,thirdaddress]{Yang Xiang \fnref{fn1}}
\ead{maxiang@ust.hk}

\fntext[fn1]{Corresponding authors.}
\address[mymainaddress]{Department of Mathematics, The Hong Kong University of Science and Technology, Hong Kong, China}
\address[secondaddress]{School of Mathematics and Statistics, Wuhan University, Wuhan, Hubei, China}
\address[thirdaddress]{HKUST Shenzhen-Hong Kong Collaborative Innovation Research Institute, Futian, Shenzhen, China}

\begin{abstract}
    Amorphization during severe plastic deformation has been observed in various crystalline materials, yet its 
    underlying mechanisms remain poorly understood. This study introduces a novel phase-field model at the mesoscale, 
    integrating elastoplastic theory with a deviatoric stress-dependent transformation strain tensor to capture 
    stress-induced amorphization. The model enables quantitative predictions of amorphous phase nucleation and 
    propagation under high stress, resolving distinctive microstructural patterns such as amorphous shear bands. 
    Simulations reveal key phenomena, including avalanche-like amorphization, grain size effects, the Hall-Petch effect, 
    and surface amorphization, consistent with experimental observations. By bridging phase-field methods with elastoplastic 
    theory, this work provides a robust framework for studying amorphization as a deformation mechanism and offers valuable 
    insights for designing materials resistant to extreme mechanical conditions.
\end{abstract}


\end{frontmatter}

\linenumbers

\section{Introduction}

Amorphization under severe plastic deformation has been widely reported in crystalline materials, including alloys, ceramics, and metallic glasses 
\cite{koikeCrystaltoamorphousTransformationNiTi1990, huangDeformationinducedAmorphizationBallmilled1999, waitzMartensiticPhaseTransformations2004, 
tsuchiyaMartensiticTransformationNanostructured2006, jiangNanocrystallizationAmorphizationNiTi2013, zhaoAmorphizationNanocrystallizationSilicon2016, 
jiangMechanismsNanocrystallizationAmorphization2017, zhangAmorphousMartensiteVTi2018, guoExtremelyHardAmorphouscrystalline2018, 
keminTextureEvolutionDeformation2020, zhaoAmorphizationExtremeDeformation2021, zhaoDirectionalAmorphizationCovalentlybonded2021, 
huaNanocompositeNiTiShape2021, liAmorphizationMechanicalDeformation2022, huaShearinducedAmorphizationNanocrystalline2022}. Two primary mechanisms 
have been proposed for this solid phase transition: strain-induced amorphization and stress-induced amorphization \cite{koikeCrystaltoamorphousTransformationNiTi1990, 
Ovidko2012NanoscaleAA, zhaoDirectionalAmorphizationCovalentlybonded2021, idrissiAmorphizationDeformationMechanism2022, 
idrissiFormationMechanismsIntragranular2022, zhaoAmorphizationmediatedPlasticity2023}. Strain-induced amorphization is attributed to 
the progressive accumulation of defects, such as dislocations and grain boundaries, which destabilize the crystalline structure over time 
\cite{koikeCrystaltoamorphousTransformationNiTi1990, Ovidko2012NanoscaleAA}. In contrast, stress-induced amorphization involves the direct nucleation 
of amorphous phases under high stress, often in the absence of significant defect activities, and is triggered by elastic instability or localized 
stress concentrations \cite{koikeRoleShearInstability1990, Tochigi2014InST, prokoshkinNanostructureFeaturesStressinduced2019,
zhaoDirectionalAmorphizationCovalentlybonded2021}. For example, stress-induced amorphization at grain boundaries, without dislocation activity, 
challenges strain-centric theories \cite{yamada_influence_1993}. Similarly, in situ deformation experiments have revealed the formation of amorphous 
shear bands in highly stressed regions, further supporting the stress-driven mechanism \cite{Tochigi2014InST}. This work focuses on stress-induced 
amorphization to investigate its role in the deformation behavior of crystalline materials and to provide new insights into the underlying mechanisms 
governing this transition.

Molecular dynamics (MD) simulations have provided atomic-scale insights into amorphization, such as simulations on nanoindentation \cite{fanGrainSizeEffects2018} 
and shear deformation \cite{huaNanocompositeNiTiShape2021}, revealing grain boundaries as nucleation sites \cite{liAmorphizationMechanicalDeformation2022}. 
However, MD simulations are limited by their reliance on picosecond-scale timescales and nanometer-scale domains, which restrict their ability to 
capture macroscale plasticity or experimentally relevant strain rates. In contrast, phase field (PF) modeling offers a mesoscale framework capable of 
simulating microstructure evolution driven by thermodynamic forces and has been successfully applied in martensitic transformation, deformation twinning, 
fracture, et al. \cite{levitasThermomechanicalTheoryMartensitic1998, artemevThreedimensionalPhaseField2001, yamanakaElastoplasticPhasefieldSimulation2008, 
finelPhaseFieldMethods2010, levitasInterfacePropagationMicrostructure2010, yamanakaElastoplasticPhasefieldSimulation2010, claytonPhaseFieldModel2011, 
heoPhasefieldModelDeformation2011, kundinPhasefieldModelIncoherent2011, yedduThreedimensionalPhasefieldModeling2012, levinPhasefieldSimulationStressinduced2013, 
levitasMultipleTwinningVariantvariant2013, sheFiniteElementSimulation2013, yedduStraininducedMartensiticTransformation2013, malikEffectExternalLoading2013, 
Schmitt2014CrystalPA, zhongPhasefieldModelingMartensitic2014, levitasInteractionPhaseTransformations2015, javanbakhtInteractionPhaseTransformations2015, 
vattrePolymorphismIronHigh2016, schmidtPhaseFieldModel2017, liuIntegratedCrystalPlasticity2018, mirzakhaniPhaseFieldelasticityAnalysis2018, xiePhaseFieldModeling2018, 
basakFiniteElementProcedure2019, pengPhaseFieldSimulation2020, xuPhaseFieldSimulation2020, xuPhaseFieldSimulation2021, levitasPhaseTransformationsFracture2021, 
maPhaseFieldModeling2021, basakMultiphasePhasefieldStudy2023, yaoCoupledPhasefieldCrystal2024}. Phase field theory holds promise for investigating deformation-induced 
amorphization \cite{Clayton2014PhaseFT, xiPhaseFieldCrystal2017, claytonPhaseFieldModeling2019, arriccaFiniteStrainContinuum2025}. For instance, 
\citet{claytonPhaseFieldModeling2019} proposed a phase field model of pressure-induced amorphization in boron carbide ceramic, incorporating elastic instability 
as a trigger to amorphization. However, existing PF models neglect critical aspects of stress-induced amorphization, such as the coupling between severe plastic 
deformation and phase transformation, and the special transformation strain for amorphization. Notably, PF studies incorporating finite-strain theory provide a 
robust framework for studying multiphase problems under severe plastic deformation \cite{levitasPhasefieldTheoryMartensitic2013, levinPhasefieldSimulationStressinduced2013, 
hongPhasefieldModelSystems2013, borukhovichLargeDeformationFramework2015, schneiderStressCalculationPhasefield2017, pengPhaseFieldSimulation2020}. What is more,
new transformation strains, which break from classical crystallographic definitions but are thermodynamically consistent, are also developed 
\cite{levitasCoherentSolidLiquid2011}. Considering these advancements on PF modeling, PF approaches can overcome the limitations of MD simulation 
and uncover the deformation mechanisms governing the formation and behavior of amorphous phases in large-deformed crystalline materials. This makes PF modeling a 
promising tool for studying deformation-induced amorphization.

We propose a novel phase field model that couples elastoplasticity with phase transformation to study stress-induced amorphization. The local crystallinity
is described by a phase variable, which evolves under a driving force derived from strain energy. Building on the work of \citet{levitasCoherentSolidLiquid2011} 
and \citet{gaoImplicitFiniteElement2006}, our model introduces a deviatoric stress-dependent transformation strain. Specifically, the volumetric strain is governed 
by the density ratio of amorphous and crystalline phases, while the deviatoric strain is proportional to the applied deviatoric stress, departing from crystallographic 
definitions (e.g., for martensite and twins). Thermodynamics consistency for the kinetic equations is also guaranteed by the Clausius-Duhem inequality 
\cite{levitasCoherentSolidLiquid2011}. The model couples plasticity in both phases through stored elastic strain energy, enabling the simulation of defect evolution, 
such as shear bands and dislocations, during amorphization. By combining transformation strain energy and plastic work, the severely deformed materials overcome 
the energy barrier of amorphization in the model. The proposed framework for studying stress-induced amorphization at the continuum scale allows for quantitative 
investigations of microstructural evolution and offers new insights into the mechanisms driving amorphization under severe deformation.

We investigate key phenomena in stress-induced amorphization, including amorphous shear bands, grain size effect, and surface amorphization. Simulations 
reveal the formation and propagation of  amorphous shear bands under severe deformation, driven by the elastic instability in crystalline phases, consistent 
with experimental observation \cite{tatyaninAmorphousShearBands1997, schuhMechanicalBehaviorAmorphous2007, huaShearinducedAmorphizationNanocrystalline2022, 
liAmorphizationMechanicalDeformation2022, idrissiFormationMechanismsIntragranular2022}. Grain size effects are also captured, showing that smaller grains 
facilitate amorphization, while larger grains increase the difficulty of forming amorphous phases, aligning with experimental findings \cite{fanGrainSizeEffects2018, 
huaNanocompositeNiTiShape2021, huaShearinducedAmorphizationNanocrystalline2022, xuGeneralizationHallPetchInverse2023}. A novel discovery is the critical-like
behavior of stress-induced amorphization, characterized by avalanche dynamics associated with the formation of amorphous shear bands. This behavior, observed 
for the first time, highlights the relationship between amorphization and plasticity. Additionally, the simulations reproduces the classic Hall-Petch effect, where 
smaller grains lead to higher yield stress, further validating the ability of the proposed model to capture fundamental material behaviors \cite{xuGeneralizationHallPetchInverse2023}. 
Finally, three-dimensional compression simulations demonstrate that amorphous phases nucleate at surfaces and stress concentrators, consistent with in situ 
TEM studies \cite{zhaoAmorphizationNanocrystallizationSilicon2016, guoExtremelyHardAmorphouscrystalline2018, huaNanocompositeNiTiShape2021, zhaoAmorphizationExtremeDeformation2021, 
huaShearinducedAmorphizationNanocrystalline2022}. These findings underscore the ability of the proposed model to quantitatively investigate the essential features 
of stress-induced amorphization.

The paper is structured as follows: Section 2 introduces the phase field model for amorphization, incorporating finite-strain elastoplasticity to 
capture the coupling between stress and phase transformation. Section 3 develops a linearized theory to improve computational efficiency, enabling 
faster simulations without compromising accuracy. Section 4 applies the model to 2D and 3D scenarios, providing insights into amorphous shear bands, 
grain size effects, and surface amorphization. Finally, Section 5 concludes the study by summarizing key findings and discussing future directions 
for extending the model to more complex deformation mechanisms.

\section{Phase field model for amorphization}

This section presents the phase field model coupled with finite strain theory to study stress-induced amorphization. The model is developed under the a
ssumption of an isothermal system, which simplifies the analysis by neglecting the effects of thermal loads associated with the two phases. This assumption 
allows us to focus on the mechanical and structural aspects of amorphization.

To describe the kinematics of finite deformation, we consider a reference configuration $\Omega_0 \subset \mathbb{R}^3$ and a material 
point $\bm{x}$ within $\Omega_0$. The deformation is represented by a mapping $\mathcal{X}(\bm{x}): \bm{x} \in \Omega_0 \rightarrow \bm{X} 
\in \Omega$, which maps the material point $\bm{x}$ in the reference configuration to its position $\bm{X}$ in the deformed configuration 
$\Omega$. The deformation gradient, referred to the undeformed configuration, is denoted by $\bm{F} = \frac{\partial \mathcal{X}}{\partial \bm{x}}$. 
This mathematical framework provides the foundation for modeling the coupling between stress and phase transformation during amorphization.

\subsection{Order parameters}

We define a continuous field variable, $\eta \in [0,1]$, to describe the crystallinity of materials. Specifically, $\eta=0$ 
represents the crystalline phase, while $\eta=1$ corresponds to the amorphous phase. The amorphous volume fraction, $V_g$, 
is described by the interpolation function:
\begin{equation}
    V_g = h(\eta) = 2\eta^2 - \eta^4, \label{interpolation}
\end{equation}
which ensures smooth transitions between phases. The derivatives of $h(\eta)$ vanish at $\eta=0$ and $\eta=1$, ensuring 
stability at the pure crystalline and amorphous states. This specific form of $h(\eta)$ is chosen for its simplicity and 
ability to capture the nonlinear evolution of the amorphous phase.

To account for the mass density of materials, we denote $\rho_c$ and $\rho_g$ as the mass densities of the crystalline and 
amorphous phases, respectively. The overall mass density of the system is expressed using the standard rule of mixtures 
\cite{arriccaFiniteStrainContinuum2025}:
\begin{equation}
    \rho = (1 - V_g)\rho_c + V_g\rho_g.
\end{equation}
This formulation provides a consistent framework for linking the phase variable $\eta$ to the physical properties, enabling 
the study of stress-induced amorphization.

\subsection{Kinematics}

\begin{figure}
    \centering
    \includegraphics[width=0.5\linewidth]{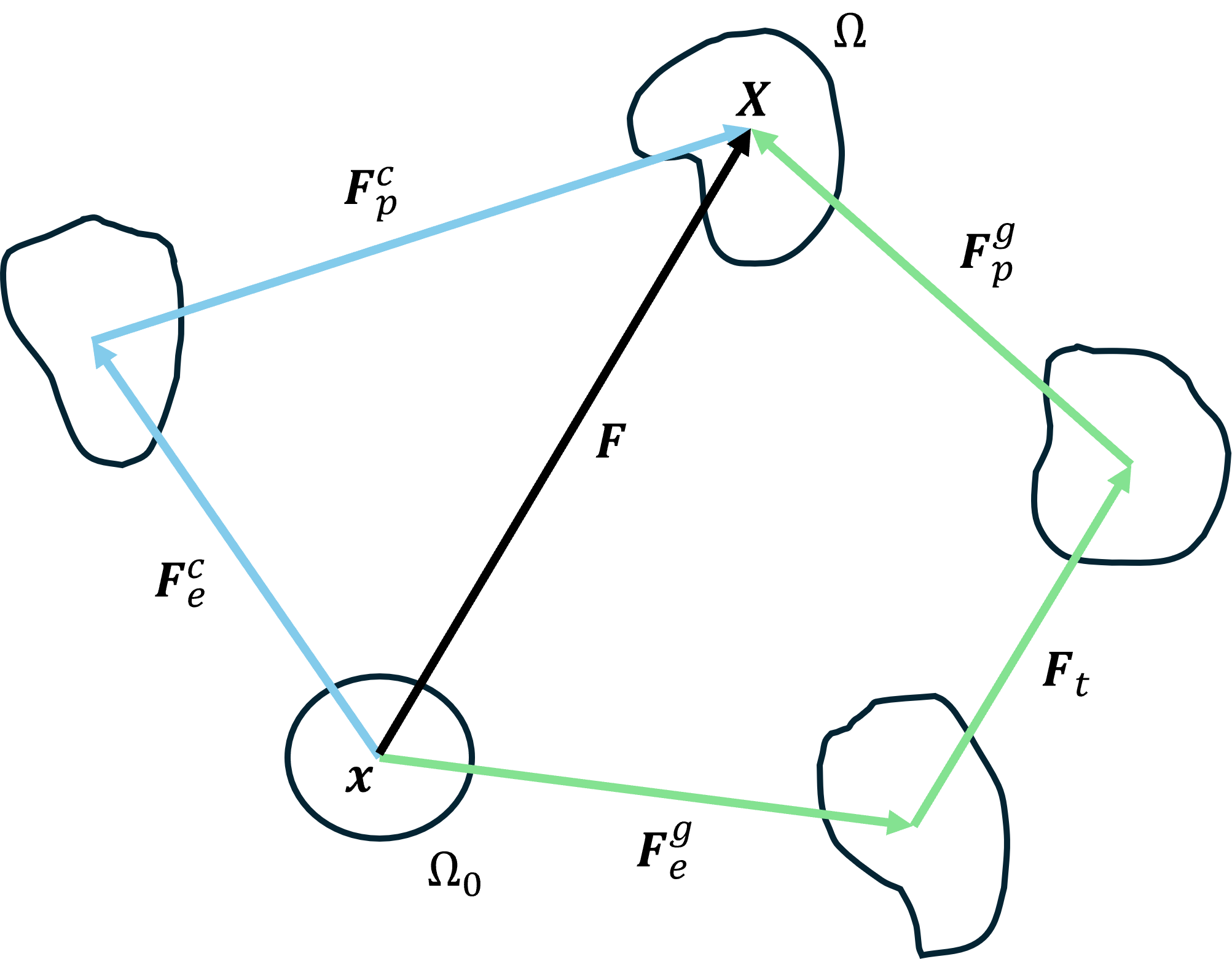}
    \caption{The decomposition of the deformation gradient for each phase.  For the crystalline phase, the deformation 
    gradient is expressed as $\bm{F}_c = \bm{F}_e^c \bm{F}_p^c$, where $\bm{F}_e^c$ and $\bm{F}_p^c$ denote the elastic 
    and plastic components, respectively. In contrast, the deformation gradient for the amorphous phase is given by $\bm{F}_g 
    = \bm{F}_e^g \bm{F}_t \bm{F}_p^g$, where $\bm{F}_t$ accounts for the transformation strain during amorphization. $\bm{F}_e^g$
    and $\bm{F}_p^g$ represent the elastic and plastic components of the amorphous phase, respectively.}
    \label{deformationgradient}
\end{figure}

The assumption of equal deformation gradients between the crystalline and amorphous phases is commonly used in multiphase modeling 
\cite{arriccaFiniteStrainContinuum2025}. This assumption simplifies the analysis by ensuring consistency in the mechanical 
response of the two phases. Specifically, we assume that the deformation gradients of the crystalline and amorphous phases 
are identical, leading to the following equality (Figure \ref{deformationgradient}) \cite{vattrePolymorphismIronHigh2016, 
schneiderStressCalculationPhasefield2017},
\begin{equation}
    \bm{F}=\bm{F}_c=\bm{F}_g,
\end{equation}
where $\bm{F}_c$ and $\bm{F}_g$ are the deformation gradients of the crystalline and amorphous phases, respectively. 

Figure \ref{deformationgradient} illustrates the decomposition of the deformation gradient for each phase, which provides 
a consistent framework for modeling the mechanical contributions of both phases during stress-induced amorphization.

For crystalline phases, the deformation gradient is multiplicatively decomposed into elastic and plastic components as,
\begin{equation}
    \bm{F}_c=\bm{F}_e^c\bm{F}_p^c,\label{totalstrainc}
\end{equation}
where $\bm{F}_e^c$ represents the elastic deformation gradient, and $\bm{F}_p^c$ accounts for the inelastic part of deformation
gradient. The evolution of the plastic deformation gradient is governed by:
\begin{equation}
    \dot{\bm{F}}_p^c=\bm{L}_p^c\bm{F}_p^c,\label{plasticflow}
\end{equation}
where $\bm{L}_p^c$ is the plastic velocity gradient. 

The plastic velocity gradient is expressed as the sum of the shear rates on $N_s$ slip systems \cite{mieheComparativeStudyStress2001a, 
levinPhasefieldSimulationStressinduced2013, borukhovichLargeDeformationFramework2015, liuIntegratedCrystalPlasticity2018, 
rotersOverviewConstitutiveLaws2010}:
\begin{equation}
    \bm{L}_p^c=\sum_{\alpha=1}^{N_s} \dot{\gamma}^{\alpha}\bm{m}^{\alpha}\otimes \bm{n}^{\alpha},\label{LPC}
\end{equation}
where $\dot{\gamma}^{\alpha}$ is the shear rate on the slip system $\alpha$, and the vectors $\bm{m}^{\alpha}$ and $\bm{n}^{\alpha}$ 
are the slip direction and slip plane normal, respectively.

An isotropic plastic flow rule is assumed for the crystalline phases. The shear rate on slip system $\alpha$ is given by
\cite{rotersOverviewConstitutiveLaws2010, hongPhasefieldModelSystems2013}:
\begin{equation}
    \dot{\gamma}^{\alpha}=\dot{\gamma}_0\left \vert \frac{\tau^{\alpha}}{\tau_c^{\alpha}}\right \vert^{\frac{1}{m}}\text{sign}(\tau^{\alpha}),
    \label{plasticrate}
\end{equation}
where $\tau^{\alpha}$ is the resolved shear stress, $\tau_c^{\alpha}$ is the slip resistance, $\dot{\gamma}_0$ is the reference shear 
rate, and $m$ is the strain rate sensitivity. The slip resistance evolves according to the hardening behavior:
\begin{equation}
    \tau_c^{\alpha}=\sum_{\beta=1}^{N_s}h_{\alpha\beta}\vert\dot{\gamma}^{\beta}\vert,
\end{equation}
where $h_{\alpha\beta}$ is the hardening matrix, and $\beta$ indexes the slip systems.

For the amorphous phase, the deformation gradient is decomposed as,
\begin{equation}
    \bm{F}_g=\bm{F}_e^g\bm{F}_t\bm{F}_p^g,
\end{equation}
where $\bm{F}_e^g$ denotes the elastic deformation gradient. $\bm{F}_t$ is the transformation 
deformation gradient, and $\bm{F}_p^g$ represents the plastic deformation gradient. The plastic behavior of amorphous phases 
is governed by the evolution of the plastic deformation gradient, 
\begin{equation}
    \dot{\bm{F}}_p^g=\bm{L}_p^g\bm{F}_p^g,
\end{equation}
where the plastic velocity gradient is expressed as \cite{gaoImplicitFiniteElement2006, 
kassnerCreepAmorphousMetals2015, ferreiraEfficientFiniteStrain2023}, 
\begin{equation}
    \bm{L}_g^p=\dot{\gamma}_g(\bm{R}_e^g)^T\bm{N}\bm{R}_e^g,
\end{equation}
with $\bm{R}_e^g=\bm{F}_e^g\left[(\bm{F}_e^g)^T\bm{F}_e^g\right]^{-\frac{1}{2}}$ as the elastic rotation 
tensor and $\bm{N}$ as the visco-plastic flow vector:
\begin{equation}
    \bm{N}=\sqrt{\frac{1}{2}}\frac{\hat{\bm{\tau}}^g}{\Vert\hat{\bm{\tau}}^g\Vert},
\end{equation}
where $\hat{\bm{\tau}}^g$ as the deviatoric second Kirchhoff stress tensor and $\Vert \cdot\Vert$ denotes the Frobenius norm. 
The visco-plastic multiplier $\dot{\gamma}_g$ is given by,
\begin{equation}
    \dot{\gamma}_g=\frac{1}{A}\sinh{\frac{\hat{\tau}_{eq}^g}{\tau^*}},\label{creep}
\end{equation}
where $\hat{\tau}_{eq}^g=\sqrt{\frac{1}{2}}\Vert\hat{\bm{\tau}}^g\Vert$ is the equivalent Kirchhoff stress tensor, $\tau^*=\frac{2k_BT}
{V_{\text{atom}}}$ is the reference stress, and $A$ is a material parameter. $V_{\text{atom}}$ is the atomic volume, $k_B$ is the Boltzmann 
constant and $T$ is the absolute temperature.

As a summary, the plastic deformation in amorphous phases evolves as follows:
\begin{equation}
    \bm{L}_g^p=\frac{1}{A}\sinh{\frac{\hat{\tau}_{eq}^g}{\tau^*}}(\bm{R}_e)^T\bm{N}\bm{R}_e.\label{LPG}
\end{equation}

Inspired by \citet{levitasCoherentSolidLiquid2011} and the creep behavior of amorphous materials \cite{kassnerCreepAmorphousMetals2015}, 
we introduce a novel transformation strain tensor to describe amorphization. Unlike martensitic transformations, this transformation strain 
is not determined by crystallography but is governed by a thermodynamically consistent kinetic equation. The transformation strain is 
decomposed as,
\begin{equation}
    \bm{F}_t=\frac{1}{3}\varepsilon_0\bm{I}+\bm{F}_t^d,
\end{equation}
where $\varepsilon_0$ represents volumetric strain due to change in mass density between the crystalline and amorphous phases,
and $\bm{F}_t^d$ is the deviatoric part of the transformation strain. The evolution of $\bm{F}_t^d$ is governed by:,
\begin{equation}
    \dot{\bm{F}}_t^d=\bm{L}_t\bm{F}_t^d,
\end{equation}
where the velocity gradient $\bm{L}_t$ is formulated as:
\begin{align}
    \bm{L}_t&=\eta(1-\eta)\dot{\eta}\text{sign}(\dot{\eta})\dot{\gamma}_t\sinh{\frac{\hat{\tau}_{eq}^g}{\tau^*}}(\bm{R}_e)^T\bm{N}\bm{R}_e\notag\\
    &=\sqrt{\frac{1}{2}}\eta(1-\eta)\dot{\eta}\text{sign}(\dot{\eta})\dot{\gamma}_t\sinh{\frac{\hat{\tau}_{eq}^g}{\tau^*}}(\bm{R}_e)^T
    \frac{\hat{\bm{\tau}}^g}{\Vert\hat{\bm{\tau}}^g\Vert}\bm{R}_e, \label{Lt}
\end{align}
which $\dot{\gamma}_t$ is the shear rate for amorphization. In the next section, we will derive the work-conjugate force of the 
transformation strain, which is the second Piola-Kirchhoff stress, using the second law of thermodynamics. Then, we combine this 
thermodynamic consistency requirement and the creep behavior of amorphous phases to define the new transformation strain and 
hence $\sinh{\frac{\hat{\tau}_{eq}^g}{\tau^*}}(\bm{R}_e)^T\bm{N}\bm{R}_e$ is introduced. Furthermore, the introduction of 
$\eta(1-\eta)\dot{\eta}\text{sign}(\dot{\eta})$ ensures that the transformation strain evolves only during phase transitions, 
maintaining stability in pure crystalline and amorphous states.

\subsection{Free energy functional}

The specific (per mass) free energy functional of the system, $\psi$, governs the evolution of the material system and consists 
of three main components: the local phase separation energy $\psi^{ch}$, the gradient energy $\psi^{\nabla}$, and the elastic 
strain energy in crystalline and amorphous phases, $\psi_e^c$ and $\psi_e^g$, respectively. The free energy functional density
is expressed as:
\begin{equation}
\begin{aligned}
    \psi\left(\eta,\bm{F}_e^c,\bm{F}_e^g,\nabla \eta\right)&= (1-h(\eta))\psi_e^c\left(\bm{F}_e^c\right)+
    h(\eta)J_g\psi_e^g(\bm{F}_e^g(\eta)) \\
    &+\psi^{ch}\left(\eta\right)+\psi^{\nabla}\left(\nabla \eta\right),\label{freeenergyfunction}
\end{aligned}
\end{equation}
where $\eta$ is the order parameter describing the crystallinity of the material, and $h(\eta)$ represents the local volume 
fraction of the amorphous phase. The term $\psi^{ch}$ captures the bulk thermodynamic properties of the system, $\psi^{\nabla}$ 
accounts for the interfacial energy between phases, and $\psi_e^c$ and $\psi_e^g$ represent the elastic strain energy densities 
in the crystalline and amorphous phases, respectively.

This free energy functional provides a comprehensive framework for modeling the interplay between phase separation, interfacial 
effects, and mechanical deformation, which are critical for understanding stress-induced amorphization.

\subsubsection{Local phase separation energy}

The local phase separation energy density, $\psi^{ch}$, captures the bulk thermodynamic properties of the systam and is represented 
by a Landau-type polynomial:
\begin{equation}
    \psi^{ch}\left(\eta\right)=\kappa \left(\frac{A}{2}\eta^2-\frac{B}{3}\eta^3+\frac{C}{4}\eta^4\right),
    \label{MP1}
\end{equation}
where the parameter $\kappa$ represents the energy difference between the crystalline and amorphous phases. The constants $A, B$, 
and $C$ determine the shape of the local phase separation energy. 

To ensure thermodynamic consistency, the free energy of the amorphous phase must be higher than that of the crystalline phase, i.e.,
$\psi^{ch}(0)=0$, \cite{liAmorphizationMechanicalDeformation2022}. This requirement imposes the following constraint:
\begin{equation}
    \frac{A}{2}-\frac{B}{3}+\frac{C}{4}> 0.
    \label{constraint1a}
\end{equation}
Additionally, the partial derivative of $\psi^{ch}$ with respect to $\eta$ must vanish at $\eta=0$ and $\eta=1$, ensuring that 
the pure crystalline and amorphous phases correspond to local energy minima. The condition leads to the constraints:
\begin{equation}
    A-B+C=0.
    \label{constraint1b}
\end{equation}
These constraints guide the selection of suitable parameters for numerical simulations, ensuring that the energy landscape is 
physically meaningful.

The driving force associated with the local phase separation energy is given by the derivative of $\psi^{ch}$ with respect to
$\eta$:
\begin{align}
    \frac{\partial \psi^{ch}}{\partial \eta}=\kappa\left(A\eta-B\eta^2+C\eta^3\right).
    \label{MP2}
\end{align}

\subsubsection{Gradient energy}

The gradient energy represents the energy associated with the interface between two phases, capturing the cost of maintaining 
a spatially varying phase field. Its density, $\psi^{\nabla}$, is expressed as:
\begin{equation}
    \psi^{\nabla}\left(\nabla \eta\right)=\frac{1}{2}\beta\vert \nabla \eta\vert ^2,
\end{equation}
where $\beta$ is a coefficient related to the interfacial energy between the crystalline and amorphous phases. This term ensures
a smooth transition between phases by penalizing sharp gradients in the phase field variable $\eta$.

The driving force associated with the gradient energy is derived from its functional derivative with respect to $\nabla \eta$ 
and is given by,
\begin{equation}
    \nabla\frac{\partial \psi^{\nabla}}{\partial \nabla \eta}=\beta \nabla^2\eta.
    \label{MP3}
\end{equation}
This expression will be used in the Clausius-Duhem inequality to derive the kinetic equations governing the evolution of 
the phase field variable $\eta$.

\subsubsection{Strain energy}

The elastic strain energy density in the crystalline phase, $\psi_e^c$, quantifies the stored energy dur to elastic deformation 
and can be written as,
\begin{equation}
    \psi_e^c\left(\bm{F}_e^c\right)=\frac{1}{2}\bm{E}_{e}^c:\mathbb{C}_c:\bm{E}_e^c,
\end{equation}
where $\bm{E}_e^c=\frac{1}{2}\left((\bm{F}_e^c)^T\bm{F}_e^c-I\right)$ is the elastic strain tensor, and $\mathbb{C}_c$ represents 
the elastic stiffness tensor for the crystalline phase. 

The stress tensors associated with the elastic strain energy are derived as follows. The first Piola-Kirchhoff stress tensor,
$\bm{P}^c=\rho_c\frac{\partial \psi_e^c}{\partial \bm{F}_c}$, i.e.,
\begin{align}
    \bm{P}^c=\rho_c\bm{F}_e^c\cdot\frac{\partial \psi_e^c}{\partial \bm{E}_e^c}\cdot (\bm{F}_p^c)^{-1},
    \label{firstPKforcec}
\end{align}
where $\rho_c$ is the mass density in crystalline domain, assumed to remain constant during deformation. 

The Cauchy stress tensor, $\bm{\sigma}^c$, and the second Piola-Kirchhoff stress tensor $\hat{\bm{\sigma}}^c$ are expressed as,
\begin{align}
    \bm{\sigma}^c&=\rho_c\bm{F}_e^c\cdot\frac{\partial \psi_e^c}{\partial \bm{E}_e^c}\cdot(\bm{F}_e^c)^T,\\
    \hat{\bm{\sigma}}^c&=\rho_c\frac{\partial \psi_e^c}{\partial \bm{E}_e^c}.
\end{align}

These stress tensors are related through the following identities: 
\begin{align*}
    \bm{P}^c &= \bm{F}_e^c \cdot \hat{\bm{\sigma}}^c \cdot (\bm{F}_p^c)^{-1}, \\
    \bm{\sigma}^c &= \bm{F}_e^c \cdot \hat{\bm{\sigma}}^c \cdot (\bm{F}_e^c)^T, \\
    \rho_c \frac{\partial \psi_e^c}{\partial \bm{F}_e^c} &= \hat{\bm{\sigma}}^c \cdot \bm{F}_e^c.
\end{align*}

For the amorphous phase, the elastic strain energy density, $\psi_e^g$, is expressed as:
\begin{equation}
    \psi_e^g\left(\bm{F}_e^g(\eta)\right)=\frac{1}{2}\bm{E}_{e}^g:\mathbb{C}_g:\bm{E}_e^g,
\end{equation}
where $\bm{E}_e^g=\frac{1}{2}\left((\bm{F}_e^g)^T\bm{F}_e^g-I\right)$ is the elastic strain tensor, and 
$\mathbb{C}_g$ represents the elastic coefficient tensor for the amorphous phase. The elastic deformation 
gradient $\bm{F}_e^g(\eta)=(\bm{F}_t(\eta)\bm{F}_p^g)^{-1}\bm{F}$ depends on the phase field variable $\eta$ 
due to the transformation strain $\bm{F}_t$. Similar to the crystalline phase, the first Piola-Kirchhoff 
stress tensor $\bm{P}^g$, the Cauchy stress tensor $\bm{\sigma}^g$, and the second Piolar-Kirchhoff stress 
tensor $\hat{\bm{\sigma}}^g$ with respect to the unloaded configuration $\Omega_0$ are given by,
\begin{equation}
\begin{aligned}
    \bm{P}^g&=\rho_g\bm{F}_e^g\cdot\frac{\partial \psi_e^g}{\partial \bm{E}_e^g}\cdot (\bm{F}_t\bm{F}_p^g)^{-1},\\
    \bm{\sigma}^g&=\rho_g\bm{F}_e^g\cdot\frac{\partial \psi_e^g}{\partial \bm{E}_e^g}\cdot(\bm{F}_e^g)^T,\\
    \hat{\bm{\sigma}}^g&=\rho_g\frac{\partial \psi_e^g}{\partial \bm{E}_e^g},\label{firstPKforceg}
\end{aligned}
\end{equation}
where the density of the amorphous phase $\rho_g$ is also assumed to remain constant during deformation. We 
also find the following equations relating stress tensors 
\begin{align*}
    \bm{P}^g &= \bm{F}_e^g \cdot \hat{\bm{\sigma}}^g \cdot (\bm{F}_t \bm{F}_p^g)^{-1}, \\
    \bm{\sigma}^g &= \bm{F}_e^g \cdot \hat{\bm{\sigma}}^g \cdot (\bm{F}_e^g)^T, \\
    \rho_g \frac{\partial \psi_e^g}{\partial \bm{F}_e^g} &= \hat{\bm{\sigma}}^g \cdot \bm{F}_e^g.
\end{align*}

The driving force associated with the elastic strain energy will be derived from the Clausius-Duhem inequality
later. 

\subsubsection{Balance laws}

During deformation, the local form of the linear momentum balance governs the mechanical equilibrium of the system and 
is expressed as::
\begin{equation}
    \nabla \cdot \bm{P}=0,
\end{equation}
where $\bm{P}$ is the first Piola-Kirchhoff stress tensor. This equation ensures that the internal and external 
forces are balanced at every material point, providing a fundamental framework for describing deformation behavior.

To model isothermal and irreversible deformation processes, it is essential to formulate constitutive relations that 
are thermodynamically consistent. This ensures that the model adheres to the second law of thermodynamics and 
accurately captures the energy dissipation associated with deformation. The next section discusses the thermodynamic 
formalism required to derive these constitutive relations and their role in the phase field model for stress-induced 
amorphization.

\subsubsection{Second law of thermodynamics}

The second law of thermodynamics provides a rigorous framework for deriving the driving forces governing the evolution
of strains and amorphization. Under isothermal conditions, the Clausius-Duhem inequality is written per unit reference volume
as \cite{levitasThermomechanicalTheoryMartensitic1998, levitasInterfacePropagationMicrostructure2010, levitasPhasefieldTheoryMartensitic2013, 
vallicottiVARIATIONALLYCONSISTENTCOMPUTATIONAL2018, levitasRecentSituExperimental2023}, 
\begin{equation}
    \int_{\Omega_0}\bm{P}:\dot{\bm{F}}-\rho_c\dot{\psi}d\Omega_0\geq 0,\label{CDinequality}
\end{equation}
where $\bm{P}$ is the first Piola-Kirchhoff stress tensor, $\bm{F}$ is the deformation gradient, and $\psi$ is the 
free energy density. The inequality ensures that the rate of energy dissipation is non-negative.

The rates of total deformation $\dot{\bm{F}}$ and free energy density $\dot{\psi}$ are given by:
\begin{align}
    \dot{\bm{F}}&=\dot{\bm{F}}_e^c\bm{F}_p^c+\bm{F}_e^c\dot{\bm{F}}_p^c,\\
    \dot{\psi}&=(1-h(\eta))\frac{\partial \psi_e^c}{\partial \bm{F}_e^c}:\dot{\bm{F}}_e^c
    +h(\eta)J_g\frac{\partial \psi_e^g}{\partial \bm{F}_e^g}:\dot{\bm{F}}_e^g+\frac{\partial \psi^{\nabla}}
    {\partial \nabla \eta}\cdot\dot{\nabla\eta}\notag\\
    &+\left[(J_g\psi_e^g-\psi_e^c)h'(\eta)+h(\eta)\rho_g\frac{\partial \psi_e^g}{\partial \eta}+\frac{\partial \psi^{ch}}{\partial \eta}\right]\dot{\eta},
\end{align}
which can be obtained from Equation \ref{totalstrainc} and \ref{freeenergyfunction}. 

Substituting these expressions into the Clausius-Duhem inequality leads to:
\begin{equation}
\begin{aligned}
    \int_{\Omega_0}&\left[\bm{F}_p^c\cdot\bm{P}-(1-h(\eta))\rho_c\frac{\partial \psi_e^c}{\partial \bm{F}_e^c}
    -h(\eta)\rho_g\bm{F}_p^c\cdot(\bm{F}_t\cdot\bm{F}_p^g)^{-1}\cdot\frac{\partial \psi_e^g}{\partial \bm{F}_e^g}\right]:\dot{\bm{F}}_e^c\\
    &+\left[\bm{P}-h(\eta)\rho_g(\bm{F}_t\cdot\bm{F}_p^g)^{-1}\cdot\frac{\partial \psi_e^g}{\partial \bm{F}_e^g}\right]:\bm{F}_e^c\cdot\dot{\bm{F}}_p^c\\
    &+h(\eta)\rho_g\bm{F}_p^g\cdot(\bm{F}_t\cdot\bm{F}_p^g)^{-1}\cdot\frac{\partial \psi_e^g}{\partial \bm{F}_e^g}:\bm{F}_e^g\cdot\dot{\bm{F}}_t\\
    &+h(\eta)\rho_g(\bm{F}_t\cdot\bm{F}_p^g)^{-1}\cdot\frac{\partial \psi_e^g}{\partial \bm{F}_e^g}:\bm{F}_e^g\cdot\bm{F}_t\cdot\dot{\bm{F}}_p^g\\
    &-\rho_c\frac{\partial \psi^{\nabla}}{\partial \nabla\eta}\cdot\dot{\nabla\eta}-\left[(\rho_g\psi_e^g-\rho_c\psi_e^c)h'(\eta)
    +h(\eta)\rho_g\frac{\partial \psi_e^g}{\partial \eta}+\rho_c\frac{\partial \psi^{ch}}{\partial \eta}\right]\dot{\eta}d\Omega_0\geq 0.
\end{aligned}
\end{equation}

The permutability of time and space differentiation in the reference configuration and the Gauss theorem are applied for the interfacial
energy term,
\begin{equation}
    \int_{\Omega_0}\frac{\partial \psi^{\nabla}}{\partial \nabla\eta}\cdot\dot{\nabla\eta}d\Omega_0=
    -\int_{\Omega_0}\nabla\frac{\partial \psi^{\nabla}}{\partial \nabla \eta}\cdot\dot{\eta}d\Omega+\int_{\Sigma_0}\bm{n}
    \cdot\frac{\partial \psi^{\nabla}}{\partial \nabla\eta}\dot{\eta}d\Sigma_0,
    \label{Gausstheorem}
\end{equation}
where an additional boundary condition for phase transitions is assumed to cancel the surface dissipation,
\begin{equation}
    \bm{n}\cdot\frac{\partial \psi^{\nabla}}{\partial \nabla\eta}=\beta\bm{n}\cdot\nabla\eta=0.
\end{equation}

Then, substituting Equation \ref{Gausstheorem} into the Clausius-Duhem inequality, the dissipation rate becomes,
\begin{equation}
\begin{aligned}
    \int_{\Omega_0}&\left[\bm{F}_p^c\cdot\bm{P}-(1-h(\eta))\rho_c\frac{\partial \psi_e^c}{\partial \bm{F}_e^c}
    -h(\eta)\rho_g\bm{F}_p^c\cdot(\bm{F}_t\bm{F}_p^g)^{-1}\cdot\frac{\partial \psi_e^g}{\partial \bm{F}_e^g}\right]:\dot{\bm{F}}_e^c\\
    &+\left[\bm{P}-h(\eta)\rho_g(\bm{F}_t\bm{F}_p^g)^{-1}\cdot\frac{\partial \psi_e^g}{\partial \bm{F}_e^g}\right]:\bm{F}_e^c\cdot\dot{\bm{F}}_p^c\\
    &+h(\eta)\rho_g\bm{F}_p^g\cdot(\bm{F}_t\bm{F}_p^g)^{-1}\cdot\frac{\partial \psi_e^g}{\partial \bm{F}_e^g}:\bm{F}_e^g\cdot\dot{\bm{F}}_t\\
    &+h(\eta)\rho_g(\bm{F}_t\bm{F}_p^g)^{-1}\cdot\frac{\partial \psi_e^g}{\partial \bm{F}_e^g}:\bm{F}_e^g\cdot\bm{F}_t\cdot\dot{\bm{F}}_p^g\\
    &-\left[(\rho_g\psi_e^g-\rho_c\psi_e^c)h'(\eta)-\rho_c\nabla\frac{\partial \psi^{\nabla}}{\partial \nabla\eta}\right.\\
    &\left.-h(\eta)\bm{F}_t^{-1}\bm{\sigma}^g:\frac{\partial \bm{F}_t}{\partial \eta}+
    \rho_c\frac{\partial \psi^{ch}}{\partial \eta}\right]\dot{\eta}d\Omega_0\geq 0.
\end{aligned}
\end{equation}

Assuming that the rate of dissipation is independent of $\dot{\bm{F}}_e^c$, the constitutive relation for the first 
Piola-Kirchhoff stress tensor is deduced as:
\begin{equation}
\begin{aligned}
    \bm{P}&=(1-h(\eta))\rho_c(\bm{F}_p^c)^{-1}\cdot\frac{\partial \psi_e^c}{\partial \bm{F}_e^c}-
    h(\eta)\rho_g(\bm{F}_t\bm{F}_p^g)^{-1}\cdot\frac{\partial \psi_e^g}{\partial \bm{F}_e^g}\\
    &=(1-h(\eta))\bm{P}^c+h(\eta)\bm{P}^g,
\end{aligned}
\end{equation}
where $\bm{P}^c$ and $\bm{P}^g$, defined in Equation \ref{firstPKforcec} and \ref{firstPKforceg}, are the stress 
tensors for the crystalline and amorphous phases, respectively. 

The dissipation rate and driving forces for $\dot{\bm{F}}_p^c$, $\dot{\bm{F}}_t$, $\dot{\bm{F}}_p^g$, and 
$\dot{\eta}$ are given as,
\begin{align}
    \mathfrak{D}:=&\bm{X}_p^c:\dot{F}_p^c+\bm{X}_t:\dot{\bm{F}}_t+\bm{X}_p^g:\dot{\bm{F}}_p^g+\dot{\bm{X}}_{\eta}\dot{\eta}\geq 0,\\
    \bm{X}_p^c&=(1-h(\eta))\hat{\bm{\sigma}}^c\cdot(\bm{F}_p^c)^{-1},\\
    \bm{X}_t&=h(\eta)\hat{\bm{\sigma}}^g\cdot(\bm{F}_t\bm{F}_p^g)^{-1},\\
    \bm{X}_p^g&=h(\eta)\bm{F}_t^{-1}\cdot\hat{\bm{\sigma}}^g\cdot(\bm{F}_t\bm{F}_p^g)^{-1},\\
    \bm{X}_{\eta}&=-\left[(J_g\psi_e^g-\psi_e^c)h'(\eta)-\nabla\frac{\partial \psi^{\nabla}}{\partial \nabla\eta}
    +\frac{\partial \psi^{ch}}{\partial \eta}\right.\\
    &\left.-h(\eta)\bm{F}_t^{-1}\bm{\sigma}^g:\frac{\partial \bm{F}_t}{\partial \eta}\right],\notag
    \label{elasticdrivingforce}
\end{align}
where $\bm{X}_p^c$, $\bm{X}_t$, $\bm{X}_p^g$, and $\bm{X}_{\eta}$ are work-conjugate driving forces for $\dot{\bm{F}}_p^c$, 
$\dot{\bm{F}}_t$, $\dot{\bm{F}}_p^g$, and $\dot{\eta}$, respectively. $\hat{\bm{\sigma}}^c$ and $\hat{\bm{\sigma}}^g$ are the elastic 
second Piola-Kirchhoff stress.

\subsection{Kinetics equations}

The evolution of the system is governed by the generalized rates of variables, which are functions of the work-conjugate driving forces.
These rates are written as:
\begin{equation}
    \bm{L}_p^c=f_p^c(\bm{X}_p^c,\eta), \bm{L}_t=f_t(\bm{X}_t,\eta),\bm{L}_p^g=f_p^g(\bm{X}_p^g,\eta), 
    \dot{\eta}=f_{\eta}(\bm{X}_{\eta},\eta).
\end{equation}
Here, $\bm{L}_p^c$, $\bm{L}_p^g$, and $\bm{L}_t$ are the plastic velocity gradients in the crystalline and amorphous phases and the 
transformational velocity gradient, respectively, as defined in Equation \ref{LPC}, \ref{LPG} and \ref{Lt}. These definitions 
ensure thermodynamical consistency with the work-conjugate forces.

The evolution of the phase variable $\eta$ is governed by the time-dependent Ginzburg-Landau (TDGL) equation, which assumes that the 
rate of change of field variables is proportional to the thermodynamic driving force:
\begin{align}
        \dot{\eta}&=M\left[-(\rho_g\psi_e^g-\rho_c\psi_e^c)h'(\eta)+\beta\nabla^2 \eta\right. \notag\\
        &\left.+h(\eta)\bm{F}_t^{-1}\bm{\sigma}^g:\frac{\partial \bm{F}_t}{\partial \eta}-
        \kappa(A\eta-B\eta^2+C\eta^3)\right],\\
        \bm{L}_p^c&=\sum_{\alpha=1}^{N_s}\dot{\gamma}^{\alpha}\bm{m}^{\alpha}\times\bm{n}^{\alpha},\\
        \bm{L}_p^g&=\frac{1}{A}\sinh{\frac{\hat{\tau}_{eq}^g}{\tau^*}}(\bm{R}_e)^T\bm{N}\bm{R}_e,\\
        \bm{L}_t&=\eta(1-\eta)\dot{\eta}\text{sign}(\dot{\eta})\dot{\gamma}_t\sinh{\frac{\hat{\tau}_{eq}^g}{\tau^*}}(\bm{R}_e)^T\bm{N}\bm{R}_e,\\
        &\bm{n}\cdot\nabla \eta=0,
        \label{MP8}
\end{align}
where the plastic velocity gradients in the crystalline and amorphous phases, as well as the transformational velocity gradient, are also
given.

These evolution equations describe the coupled deformation and amorphization processes in the system. The TDGL equation governs the evolution 
of the phase field variable $\eta$, capturing the transition between crystalline and amorphous phases. The plastic velocity gradients 
$\bm{L}_p^c$ and $\bm{L}_p^g$ describe the plastic deformation in the crystalline and amorphous phases, while $\bm{L}_t$ accounts for the 
transformation deformation gradient during amorphization. Together, these equations provide a comprehensive framework for modeling stress-induced 
amorphization and its interplay with plastic deformation.

\section{Geometric linearization}

The phase field model presented in the previous section incorporates finite strain theory to capture the coupling between stress and phase 
transformation during amorphization. While this framework provides a rigorous and comprehensive description of the deformation behavior, the 
nonlinear equations arising from finite strain mechanics significantly increase computational cost, making large-scale simulations time-consuming.

To address this challenge, we introduce a linearized theory based on linear elasticity. This simplification reduces computational complexity 
while retaining the essential physics of the model, enabling efficient simulations of stress-induced amorphization. By approximating the deformation 
behavior with linear elasticity, the model becomes more suitable for studying large-scale systems and exploring a wider range of material behaviors.

\subsection{Kinematics}

\begin{figure}
    \centering
    \includegraphics[width=0.5\linewidth]{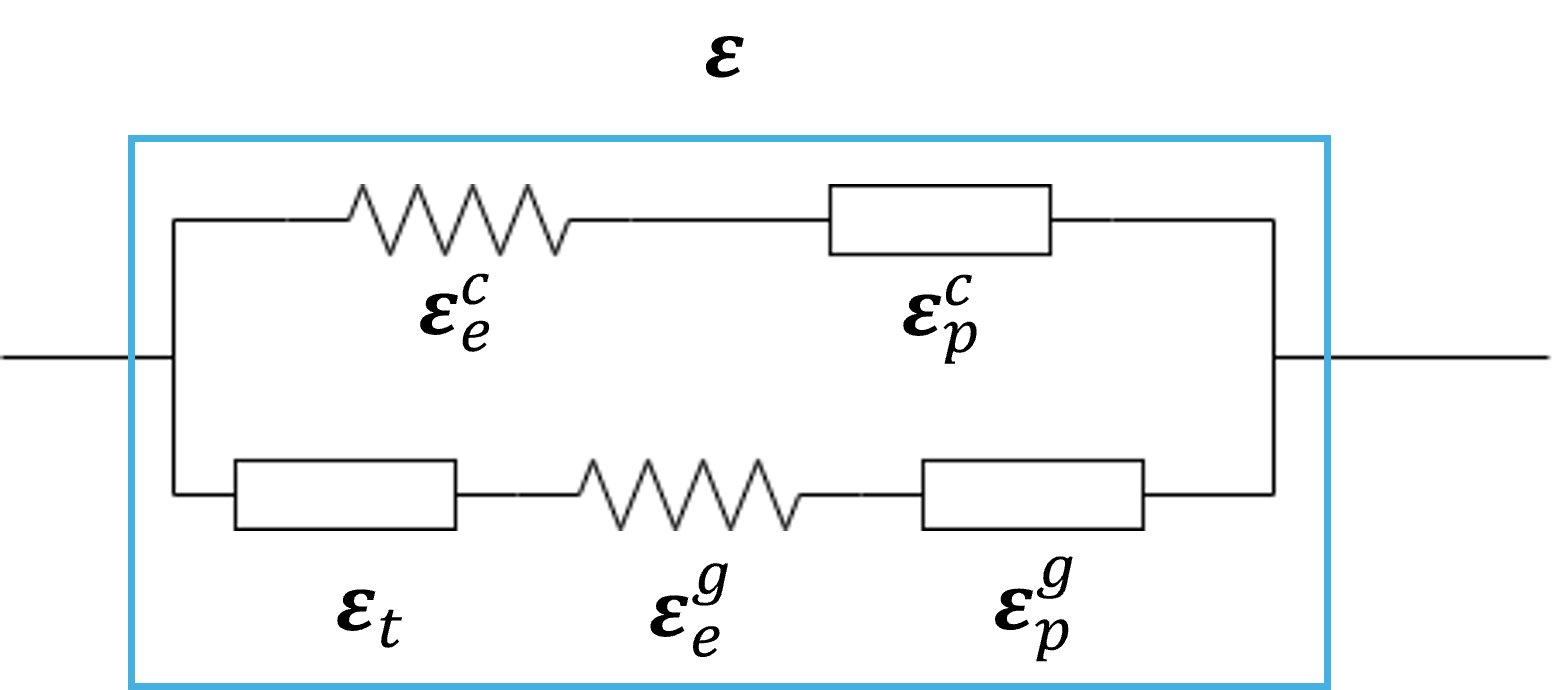}
    \caption{Schematic of the additive decomposition of the total strain for materials under deformation.}
    \label{Additiondecomposition}
\end{figure}

Under the assumption of linear elasticity, the total strain tensor of the system is written as,
$$\bm{\varepsilon}=\frac{1}{2}(\nabla\bm{u}+\nabla \bm{u}^T),$$
where $\bm{u}$ is the displacement field during deformation. For simplicity, we assume that 
the total strain is equal in both crystalline and amorphous phases \cite{arriccaFiniteStrainContinuum2025}, 
\begin{equation}
    \bm{\varepsilon}=\bm{\varepsilon}_c=\bm{\varepsilon}_g,
    \label{strainrelation}
\end{equation}
where $\bm{\varepsilon}_c$ and $\bm{\varepsilon}_g$ are the total strain tensors in the crystalline and amorphous phases, respectively. 
Figure \ref{Additiondecomposition} illustrates the additive decomposition of the total strain for material under deformation. 

In the crystalline phase, an additive decomposition is assumed for the total strain tensor, 
$$\bm{\varepsilon}_c=\bm{\varepsilon}_e^c+\bm{\varepsilon}_p^c,$$ 
where $\bm{\varepsilon}_e^c$ and $\bm{\varepsilon}_p^c$ represent the elastic and plastic strains, respectively. The elastic 
strain energy density in the crystalline phases, $\psi_e^c$, is given by:
\begin{equation}
    \psi_e^c\left(\bm{\varepsilon}_e^c\right)=\frac{1}{2}\bm{\varepsilon}_e^c:\mathbb{C}_c :\bm{\varepsilon}_e^c,
\end{equation}
where $\mathbb{C}_c$ is the elastic stiffness tensor for the crystalline phase, and the stress tensor is,
$\bm{\sigma}^c=\mathbb{C}_c:\bm{\varepsilon}_e^c$. 

In the amorphous phase, the total strain tensor is decomposed into elastic, plastic, and transformation components:
\begin{equation}
    \bm{\varepsilon}_g=\bm{\varepsilon}_e^g+\bm{\varepsilon}_t(\eta)+\bm{\varepsilon}_p^g,
\end{equation}
where $\bm{\varepsilon}_e^g$, $\bm{\varepsilon}_t$, and $\bm{\varepsilon}_p^g$ represent the elastic strain, transformation strain, 
and plastic strain, respectively. The elastic strain energy density in the amorphous phase is:
\begin{equation}
    \psi_e^g\left(\bm{\varepsilon}_e^g(\eta)\right)=\frac{1}{2}\bm{\varepsilon}_e^g:\mathbb{C}_g :\bm{\varepsilon}_e^g,
\end{equation}
where $\mathbb{C}_g$ is the elastic constant tensor of the amorphous phase, and $\bm{\sigma}^g=\mathbb{C}_g:\bm{\varepsilon}_e^g$ is the
stress tensor.

The transformation strain $\bm{\varepsilon}_t$ accounts for the volumetric and deviatoric strain contributions during amorphization.
It is defined as:
\begin{equation}
    \bm{\varepsilon}_t=\frac{1}{3}\varepsilon_0\bm{I}+\bm{e}_t,
\end{equation}
where $\varepsilon_0$ is the volumetric strain due to changes in mass density between the crystalline and amorphou phases, and $\bm{e}_t$ is 
the deviatoric transformation strain. The evolution of $\bm{e}_t$ is governed by:
\begin{equation}
    \dot{\bm{e}}_t=\eta(1-\eta)\dot{\eta}\text{sign}(\dot{\eta})\dot{\gamma}_t\sinh{\frac{\sigma_e}{\sigma_c}}\frac{\bm{S}}{\sigma_e},
\end{equation}
where $\bm{S}$ is the deviatoric stress tensor, i.e., $S_{ij}=\sigma_{ij}-\frac{1}{3}\sigma_{kk}\delta_{ij}$. Its $J_2$ 
invariant is the Mises stress, $\sigma_{e}=\sqrt{\frac{3}{2}\bm{S^T}\bm{S}}$. $\dot{\gamma}_t$ is the 
rate of transformation strain under stress. As previous discussion, the transformation strain ensures thermodynamic consistency and 
captures the stress-driven nature of amorphization, combining the creep behavior of the amorphous phase.

The plastic strain in the amorphous phase, $\bm{\varepsilon}_p^g$, follows a viscous flow model as \cite{gaoImplicitFiniteElement2006, 
kassnerCreepAmorphousMetals2015},
\begin{equation}
    \dot{\bm{\varepsilon}}_p^g=\frac{1}{A}\sinh{\frac{\sigma_{e}}{\tau^*}}\frac{\bm{S}}{\sigma_{e}},
\end{equation}
where $A$ and $\tau^*$ are material constants, and $\sigma_e$ is the equivalent stress. This model describes the time-dependent 
plastic deformation in the amorphous phase under linearized elasticity.

\subsection{Thermodynamic laws}

To derive the driving forces governing strains and amorphization, we begin by defining the specific (per mass) Helmholtz free 
energy functional as,
\begin{equation}
\begin{aligned}
    \psi(\eta,\nabla \eta, \bm{\varepsilon}_e^c,\bm{\varepsilon}_e^g)&=(1-h(\eta))\psi_c^e(\bm{\varepsilon}_e^c)
    +h(\eta)J_g\psi_g^e(\bm{\varepsilon}_e^g(\eta))\\
    &+\psi^{ch}(\eta)+\psi^{\nabla}(\nabla \eta),
    \label{linearfreeenergy}
\end{aligned}
\end{equation}
where $h(\eta)$ is the local volume fraction of the amorphous phase, and $J_g=\frac{\rho_g}{\rho_c}$ 
is the density ratio between the amorphous ($\rho_g$) and crystalline ($\rho_c$) phases.

Under isothermal conditions, the second law of thermodynamics is expressed in the form of the Clausius-Duhem inequality:
\begin{equation}
    \rho_c\mathfrak{D}=\int_{\omega_0}\bm{\sigma}:\dot{\bm{\varepsilon}}-\rho_c\dot{\psi}d\Omega_0\geq 0,
\end{equation}
where $\mathfrak{D}$ is the rate of dissipation per unit mass, $\bm{\sigma}$ is the Cauchy stress tensor, and $\dot{\psi}$ is the 
rate of change of the free energy density. Assuming constant densities $\rho_c$ and
$\rho_g$, the rate of free energy density is given by,
\begin{equation}
\begin{aligned}
    \dot{\psi}&=(1-h(\eta))\frac{\partial \psi_c^e}{\partial \bm{\varepsilon}_e^c}:\dot{\bm{\varepsilon}}_e^c+
    h(\eta)J_g\frac{\partial \psi_g^e}{\partial \bm{\varepsilon}_e^g}:\dot{\bm{\varepsilon}}_e^g+
    (J_g\psi_e^g-\psi_e^c)h'(\eta)\dot{\eta}\\
    &+h(\eta)J_g\frac{\partial \psi_e^g}{\partial \eta}\dot{\eta}+\frac{\partial \psi^{ch}}{\partial \eta}\dot{\eta}+
    \frac{\partial \psi^{\nabla}}{\partial \nabla\eta}\nabla\dot{\eta}.
\end{aligned}
\end{equation}

Substituting this expression into the Clausius-Duhem inequality and simplifying yields,
\begin{equation}
\begin{aligned}
    &\int_{\Omega_0}\left\{\bm{\sigma}-\left[(1-h(\eta))\rho_c\frac{\partial \psi_c^e}{\partial \bm{\varepsilon}_e^c}+
    h(\eta)\rho_g\frac{\partial \psi_g^e}{\partial \bm{\varepsilon}_e^c}\right]\right\}:\dot{\bm{\varepsilon}}_e^c
    +(\bm{\sigma}-h(\eta)\rho_g\frac{\partial \psi_g^e}{\partial \bm{\varepsilon}_p^c}):\dot{\bm{\varepsilon}}_p^c\\
    &-h(\eta)\rho_g\frac{\partial \psi_g^e}{\partial \bm{\varepsilon}_p^g}:\dot{\bm{\varepsilon}}_p^g
    -h(\eta)\rho_g\frac{\partial \psi_g^e}{\partial \bm{\varepsilon}_t}:\dot{\bm{\varepsilon}}_t
    -\left\{(\rho_g\psi_g^e-\rho_c\psi_c^e)h'(\eta)\right.\\
    &\left.-h(\eta)\rho_g\eta(1-\eta)\dot{\gamma}_t\sinh{\frac{\sigma_e}{\sigma_c}}
    \bm{\sigma}^g:\frac{\bm{S}}{\sigma_e}\text{sign}{(\dot{\eta})}+\rho_c\frac{\partial \psi^{ch}}{\partial \eta}-\rho_c
    \nabla\frac{\partial \psi^{\nabla}}{\partial \nabla \eta}\right\}\dot{\eta}d\Omega_0\geq 0.
\end{aligned}
\end{equation}

Assuming that the rate of dissipation is independent of $\dot{\bm{\varepsilon}}_e^c$, the constitutive relation for 
the Cauchy stress tensor can be deduced as,
\begin{equation}
    \bm{\sigma}=(1-h(\eta))\bm{\sigma}^c+h(\eta)\bm{\sigma}^g, \label{stress}
\end{equation}
where $\bm{\sigma}^c$ and $\bm{\sigma}^g$ are the Cauchy stress tensors for the crystalline and amorphous phases, respectively.

The local dissipation rate is written as,
\begin{equation*}
\begin{aligned}
    \mathfrak{D}=&\bm{X}_p^c:\dot{\bm{\varepsilon}}_p^c+\bm{X}_p^g:\dot{\bm{\varepsilon}}_p^g+\bm{X}_t:\dot{\bm{\varepsilon}}_t
    +\bm{X}_{\eta}\dot{\eta}\geq 0,\\
    \bm{X}_p^c=&\bm{\sigma}-h(\eta)\rho_g\frac{\partial \psi_g^e}{\partial \bm{\varepsilon}_p^c},
    \bm{X}_p^g=-h(\eta)\rho_g\frac{\partial \psi_g^e}{\partial \bm{\varepsilon}_p^g},\\
    \bm{X}_t=&-h(\eta)\rho_g\frac{\partial \psi_g^e}{\partial \bm{\varepsilon}_t},\\
    \bm{X}_{\eta}=&-\left[(\rho_g\psi_g^e-\rho_c\psi_e^c)h'(\eta)+\rho_c\frac{\partial\psi^{ch}}{\partial \eta}-\rho_c
    \nabla\frac{\partial \psi^{\nabla}}{\partial \nabla\eta}\right.\\
    &\left.-h(\eta)J_g\eta(1-\eta)\dot{\gamma}_t\sinh{\frac{\sigma_e}{\sigma_c}}
    \bm{\sigma}^g:\frac{\bm{S}}{\sigma_e}\text{sign}{(\dot{\eta})}\right],
\end{aligned}
\end{equation*}
where $\rho_g\frac{\partial \psi_g^e}{\partial \bm{\varepsilon}_p^c}=\bm{\sigma}^g$ and $\rho_g\frac{\partial \psi_g^e}
{\partial \bm{\varepsilon}_p^g}=\rho_g\frac{\partial \psi_g^e}{\partial \bm{\varepsilon}_t}=-\bm{\sigma}^g$. Then, the 
work-conjugate driving forces are obtained:
\begin{equation}
\begin{aligned}
    \bm{X}_p^c&=(1-h(\eta))\bm{\sigma}^c, \bm{X}_p^g=\bm{X}_t=h(\eta)\bm{\sigma}^g,\\
    \bm{X}_{\eta}&=(\psi_e^c-J_g\psi_g^e)h'(\eta)-\frac{\partial \psi^{ch}}{\partial \eta}+
    \nabla\frac{\partial \psi^{\nabla}}{\partial \nabla\eta}\\
    &+h(\eta)J_g\eta(1-\eta)\dot{\gamma}_t\sinh{\frac{\sigma_e}{\sigma_c}}
    \bm{\sigma}^g:\frac{\bm{S}}{\sigma_e}\text{sign}{(\dot{\eta})}.
\end{aligned}
\end{equation}

The momentum equilibrium equation is derived from the elastic energy as,
\begin{equation}
    \nabla\cdot \bm{\sigma}=0,
    \label{stressequation}
\end{equation}
where $\bm{\sigma}$ is the Cauchy stress tensor, defined as Equation \ref{stress}. Appropriate boundary conditions 
can be applied depending on the specific scenario.

\subsection{Kinetic equations}

The evolution of the system is governed by kinetic equations that describe the rates of change of plastic strains, transformation
strain, and the phase field variable $\eta$. These rates are functions of the work-conjugate thermodynamic forces, ensuring Thermodynamic
consistency. The general form of the kinetic equations is expressed as:
\begin{equation}
    \dot{\bm{\varepsilon}}_p^c=f_p^c(\bm{X}_p^c,\eta), \dot{\bm{\varepsilon}}_p^g=
    f_p^g(\bm{X}_p^g,\eta),\dot{\bm{\varepsilon}}_t=
    f_t(\bm{X}_t,\eta), \dot{\eta}=f_{\eta}(\bm{X}_{\eta},\eta).
\end{equation}

Using the expressions for the thermodynamic forces and the plastic flow rules, the specific kinetic equations for the system are
derived:
\begin{align}
    \dot{\eta}&=M\left[(\psi_e^c-J_g\psi_g^e)h'(\eta)-\frac{\partial \psi^{ch}}{\partial \eta}+
    \nabla\frac{\partial \psi^{\nabla}}{\partial \nabla\eta}\right.\notag\\
    &\left.+h(\eta)J_g\eta(1-\eta)\dot{\gamma}_t\sinh{\frac{\sigma_e}{\sigma_c}}
    \bm{\sigma}^g:\frac{\bm{S}}{\sigma_e}\text{sign}{(\dot{\eta})}\right], \label{phasefield0} \\
    \label{phasefield1}
    \dot{\bm{\varepsilon}}_p^c&=\sum_{\alpha=1}^N\dot{\gamma}^{\alpha}\bm{m}^{\alpha}\otimes\bm{n}^{\alpha}, \\
    \label{phasefield2}
    \dot{\bm{\varepsilon}}_p^g&=\dot{\gamma}_g\sinh{\frac{\sigma_e}{\sigma_0}}\frac{\bm{S}}{\sigma_e},\\
    \label{epsilontAmorphous}
    \dot{\bm{\varepsilon}}_t&=\dot{\bm{e}}_t=\eta(1-\eta)\dot{\eta}\text{sign}(\dot{\eta})\dot{\gamma}_t\sinh{\frac{\sigma_e}{\sigma_c}}
    \frac{\bm{S}}{\sigma_e},\\
    &\bm{n}\cdot\nabla \eta=0.\label{phasefield4}
\end{align}

Equation \ref{phasefield0} defines the evolution of the field variable $\eta$, governed by the TDGL equation.

Equation \ref{phasefield1} describes the plastic deformation in the crystalline phase, where $\alpha$ indicates a slip system, $\bm{m}^{\alpha}$ 
and $\bm{n}^{\alpha}$ are the slip direction and normal vector to the slip plane, respectively. The shear rate $\dot{\gamma}^{\alpha}$ on the slip plane reads,
$$\dot{\gamma}^{\alpha}=\dot{\gamma}_0\vert\frac{\tau^{\alpha}}{\tau_c^{\alpha}}\vert^{\frac{1}{m}}sgn(\tau^{\alpha}),$$
where $\tau^{\alpha}$ is the resolved shear stress, $\tau_c^{\alpha}$ is the slip resistance, and $m$ is the strain rate sensitivity. 

Equation \ref{phasefield2} governs the evolution of plastic strain in the amorphous phase, where $\bm{S}$ is the deviatoric stress tensor, 
$\sigma_e$ is the equivalent Mises stress, $\sigma_0$ is the yield stress, and $\dot{\gamma}_g$ is the viscoplastic multiplier.

Equation \ref{epsilontAmorphous} gives the transformation strain rate during amorphization, where $\dot{\gamma}_t$ is the reference rate of 
transformation strain, and $\sigma_c$ is the critical stress for amorphization.

The last Equation \ref{phasefield4} is the Neumann boundary condition for the phase field variable $\eta$, which ensures stability.

\subsection{Non-dimensional linearized equations}

Before performing numerical simulations, it is essential to derive dimensionless equations to simplify the computational framework and 
eliminate unnecessary parameters\cite{zhongPhasefieldModelingMartensitic2014}. 

We define the dimensionless spatial coordinates and time as:
\begin{align*}
    \tilde{x}&=\frac{x}{l_0}, \quad \tilde{y}=\frac{y}{l_0}, \quad \tilde{z}=\frac{z}{l_0},\\
    \tilde{t}&=tM\kappa,
\end{align*}
where $l_0$ is the characterized length, $\kappa$ is the energy barrier coefficient, and $M$ is the mobility of the field variable. 
The dimensionless Laplace operator is expressed as,
$$\tilde{\nabla}^2=\frac{\partial^2}{\partial \tilde{x}^2}+\frac{\partial^2}{\partial \tilde{y}^2}+
\frac{\partial^2}{\partial \tilde{z}^2}.$$
The dimensionless elastic constants and stress tensors are defined as,
\begin{align*}
    \tilde{\mathbb{C}}_c&=\frac{\mathbb{C}_c}{\kappa}, \quad \tilde{\mathbb{C}}_g=\frac{\mathbb{C}_g}{\kappa}, \\
    \tilde{\bm{\sigma}}^c&=\tilde{\mathbb{C}}_c:\bm{\varepsilon}_e^c, \quad \tilde{\bm{\sigma}}^g=
    \tilde{\mathbb{C}}_g:\bm{\varepsilon}_e^g.
\end{align*}

With dimensionless variables, the kinetic equations for the system are rewritten as:
\begin{equation}
    \begin{aligned}
        \frac{\partial \eta}{\partial \tilde{t}}&=(\tilde{\psi}_e^c-J_g\tilde{\psi}_e^g)h'(\eta)-(A\eta-B\eta^2+C\eta^3)+\tilde{\beta}
        \tilde{\nabla}^2\eta\\
        &+h(\eta)J_g\eta(1-\eta)\dot{\gamma}_t\sinh{\frac{\sigma_e}{\sigma_c}}
        \tilde{\bm{\sigma}}^g:\frac{\bm{S}}{\sigma_e}\text{sign}{(\dot{\eta})},\\
        \frac{\partial \bm{\varepsilon}_p^c}{\partial \tilde{t}}&=\sum_{\alpha=1}^N\dot{\gamma}^{\alpha}\bm{m}^{\alpha}\otimes\bm{n}^{\alpha}, \\
        \frac{\partial \bm{\varepsilon}_p^g}{\partial \tilde{t}}&=\dot{\gamma}_g\sinh{\frac{\sigma_e}{\sigma_0}}\frac{\bm{S}}{\sigma_e},\\
        \frac{\partial \bm{\varepsilon}_t}{\partial \tilde{t}}&=\eta(1-\eta)\dot{\eta}\text{sign}(\dot{\eta})\dot{\gamma}_t
        \sinh{\frac{\sigma_e}{\sigma_c}}\frac{\bm{S}}{\sigma_e},\\
        &\bm{n}\cdot\tilde{\nabla}\eta=0,
        \label{dimensionlesspf}
    \end{aligned}
\end{equation}
where $\tilde{\beta}=\frac{\beta}{\kappa l_0^2}$.

The dimensionless equations simplify the computational framework by reducing the number of parameters, making the model more efficient 
for numerical simulations. By scaling the spatial and temporal variables, the equations retain their physical accuracy while enabling 
simulations across a wide range of conditions. This approach aligns with the main goal of Section 3, which is to improve computational 
efficiency without compromising the essential physics of stress-induced amorphization.

\section{Case Study}

Numerical simulations provides insights into the microstructural evolution and stress-driven phase transformations that are challenging 
to capture experimentally. In this section, we apply the proposed phase-field model to investigate three key phenomena associated 
with stress-induced amorphization: the formation of amorphous shear bands, the effect of grain size on amorphization, and surface 
amorphization under compression. These case studies not only validate the model but also offer quantitative predictions and new 
insights into the mechanisms driving amorphization.

To perform these simulations, we utilize a combination of numerical methods tailored to the dimensionality and complexity of problems. 
The Euler method is employed for time integration, while the finite difference method is used for two-dimensional simulations. For 
three-dimensional problems, we adopt the Multiphysics Object-Oriented Simulation Environment (MOOSE) framework \cite{lindsay2022moose}, 
which provides an efficient finite element implementation for modeling the amorphization process under compression. Together, these cases 
enable a comprehensive exploration of stress-induced amorphization across different scenarios.

\subsection{Case 1: Amorphous shear band}

\begin{figure}
    \centering
    \includegraphics[width=0.5\textwidth]{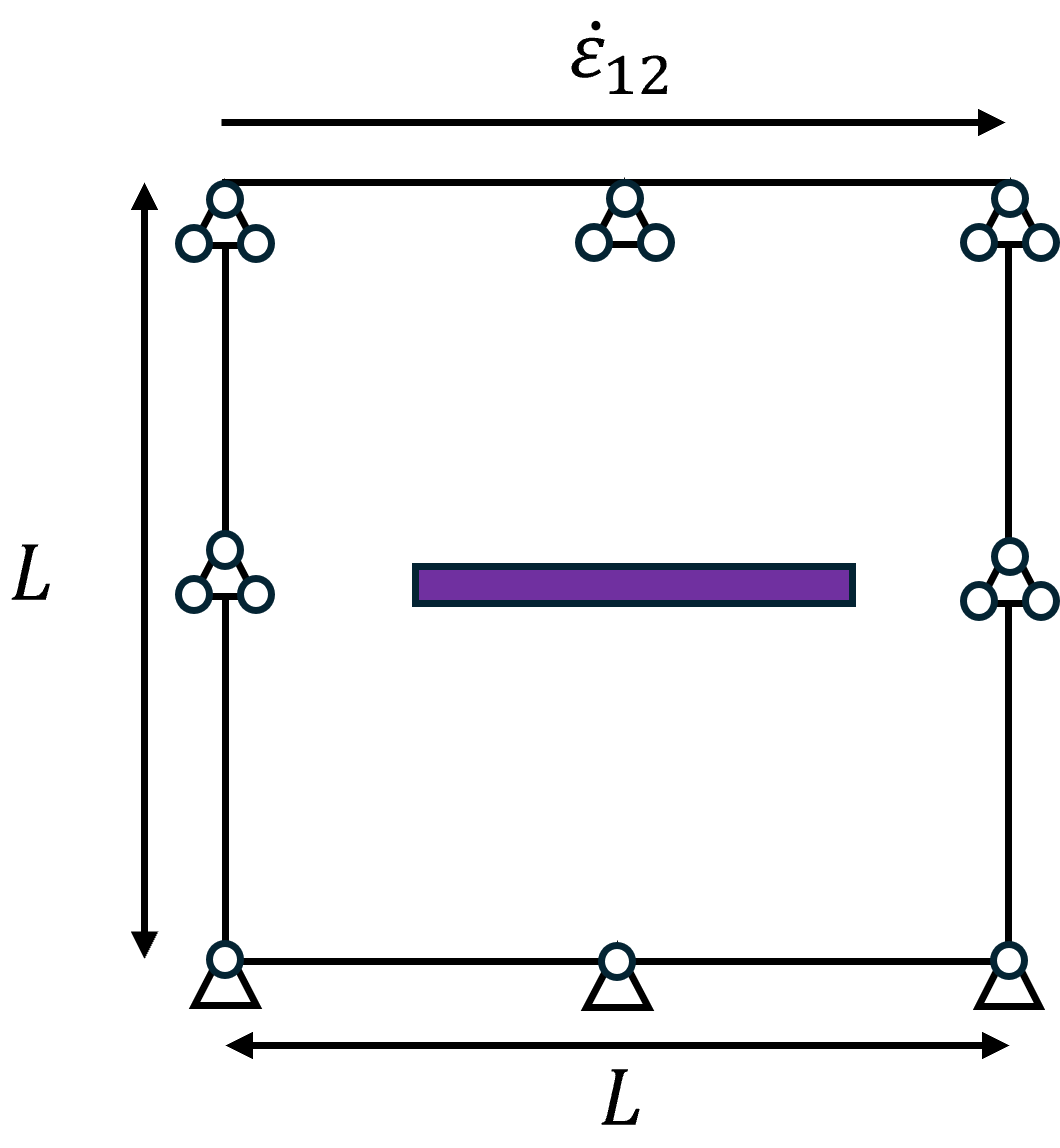}
    \caption{Numerical setup of shear in two dimensions. $\dot{\varepsilon}$ is the loading rate. $L$ is the size of the simulation 
    square cell. The purple box gives the initial amorphous defects.}
    \label{2Dsimulationset}
\end{figure}

To investigate the formation of amorphous shear bands, we simulate a 2D square cell of nanocrystalline NiTi alloy with an initial 
amorphous defect. This defect mimics microstructural imperfections, such as micro-cracks or localized disorder, commonly found in real 
materials. The simulation domain has a dimensionless length of $\tilde{L}=64$, and the amorphous defect is initialized as a rectangular 
region within the crystalline matrix (Figure \ref*{2Dsimulationset}). A pure shear deformation is applied, with periodic boundary conditions 
on the left and right boundaries and displacement-controlled conditions on the top and bottom boundaries. This setup allows us to observe how 
the amorphous phase propagates and how shear bands form under deformation.

The material parameters used in the simulation are based on previous studies \cite{schuhMechanicalBehaviorAmorphous2007, zhongPhasefieldModelingMartensitic2014, 
zhaoDirectionalAmorphizationCovalentlybonded2021, liAmorphizationMechanicalDeformation2022}. Mechanical parameters for the crystalline and amorphous 
phases of the NiTi alloy include $E_c=64.3\,GPa$, $\nu_c=0.43$, $\dot{\gamma}_0=1\times 10^{-6}$, $m=3$, $h=3\,GPa$, $\tau_{c,0}=2.7\,GPa$, $E_g=47\,GPa$, 
$\nu_g=0.3$, $\dot{\gamma}_t=5\times 10^{-3}$, $\sigma_c=1.2\,GPa$, $\dot{\gamma}_g=10^{-6}$, $\sigma_0=3.2\,GPa$. Other key parameters are 
$\kappa=4.403\times 10^7\,J\cdot m^{-3}$, $J_g=1.1$, $\varepsilon_0=0.1$, $A=12$, $B=30$, and $C=14$, $\tilde{\beta}=2$, and $\frac{d\varepsilon_{12}}{d\tilde{t}}=0.6$. 
The characterized size $l_0$ is determined using the interfacial energy density $\gamma=187\, mJ\cdot m^{-2}$, yielding $l_0=\frac{3\gamma}
{4\kappa\sqrt{2\tilde{\beta}}}=1.6\, \text{nm}$ \cite{zhongPhasefieldModelingMartensitic2014}. The simulation domain is $\tilde{L}\times 
\tilde{L}=64\times 64$, corresponding to a physical size of $L\sim 100\, \text{nm}$.

\begin{figure}[htbp]
    \centering
    \includegraphics[width=\textwidth]{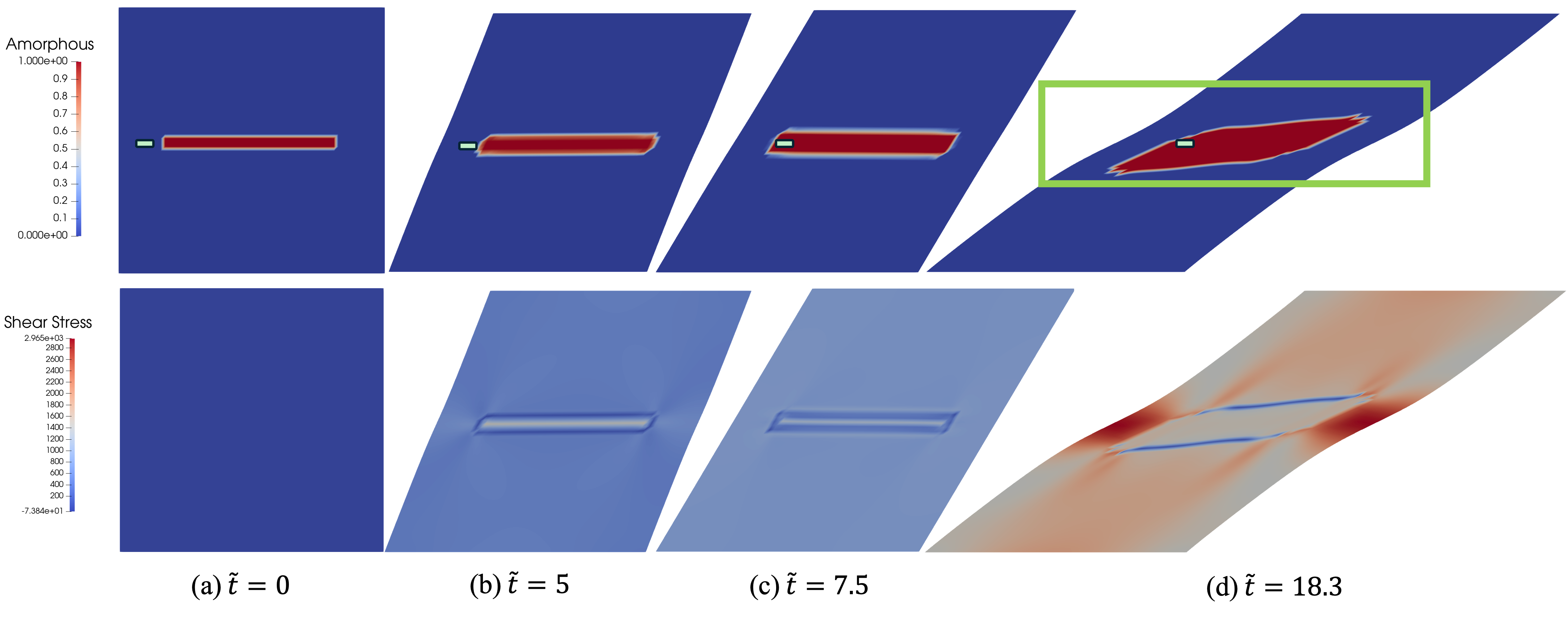}
    \caption{Results of shear deformation in two dimensions. The first row shows the evolution of the amorphous phase, while the second row presents
    the corresponding shear stress $\sigma_{xy}$. The orange box highlights the formation the amorphous shear band. The change of elastic strain
    energy in the crystalline and amorphous phases of the small green box is compared in the Figure \ref{energycomparison}.}
    \label{shearresults1}
\end{figure}

Figure \ref*{shearresults1} presents the simulation results for shear deformation. This first row shows the evolution of phase field variable $\eta$, 
and the second one presents changes of the shear stress. Initially, the NiTi alloy contains a localized amorphous defect within the crystalline 
matrix (Figure \ref{shearresults1}(a)). Upon applying shear deformation, the shear stress increase, particularly near the amorphous defect, driving 
the propagation of the amorphous phase along high-stress trajectories (Figure \ref*{shearresults1}(b) and (c)). Eventually,
a fully developed amorphous shear band emerges, characterized by highly localized strains (Figure \ref*{shearresults1}(d)).

\begin{figure}[htbp]
    \centering
    \includegraphics[width=\textwidth]{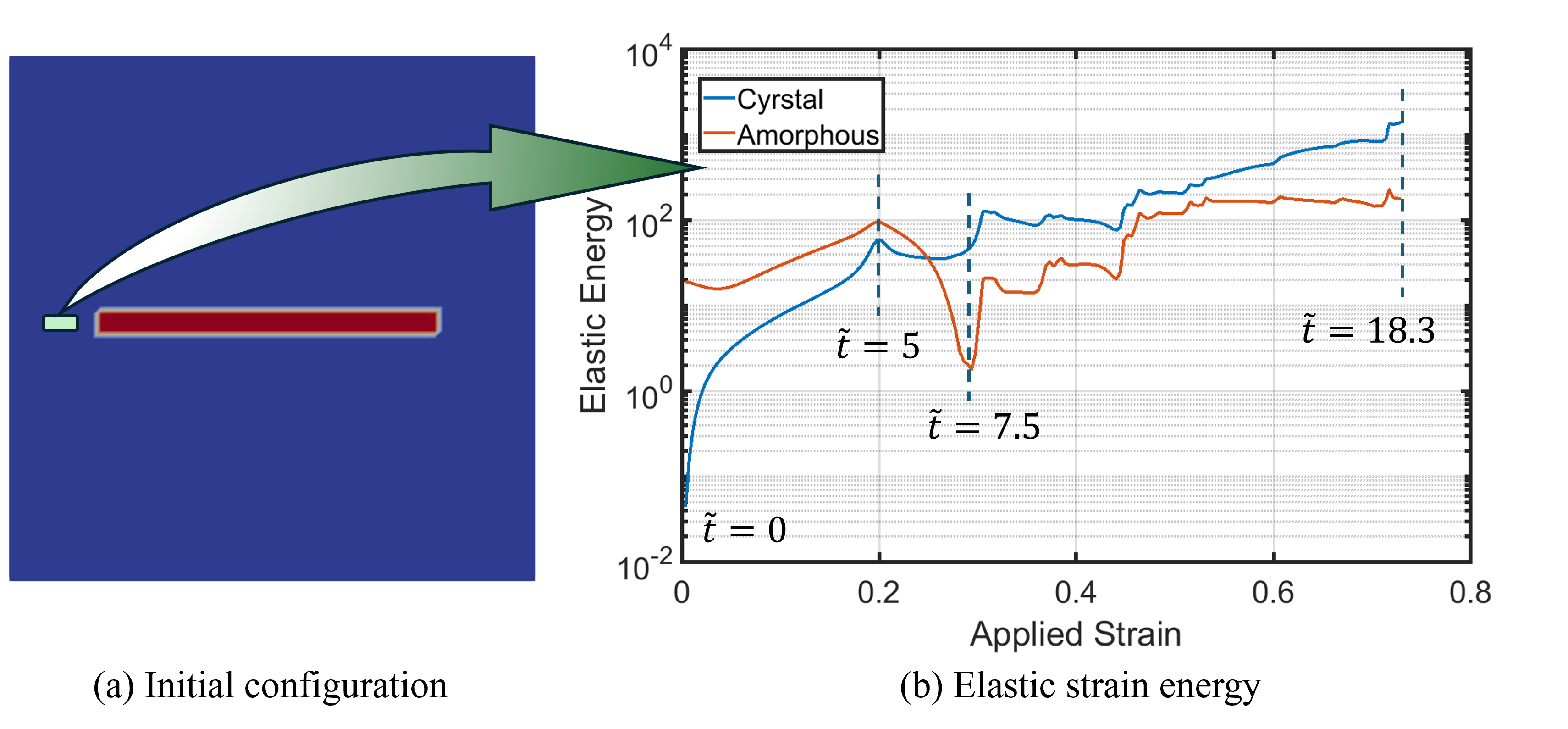}
    \caption{Comparison of the elastic strain energy stored in the crystalline and amorphous phases. (a) The small green box in Figure 
    \ref{shearresults1} is chosen for the energy comparison. (b) The elastic energy stored in the crystalline (blue) and amorphous (red) phases, 
    respectively.}
    \label{energycomparison}
\end{figure}

To further understand the role of deformation in amorphization, we examine the evolution of the elastic strain energy in the small green region (Figure 
\ref{energycomparison}(a), same as Figure \ref{shearresults1}). This region is initialized as the crystalline phase and transforms into the amorphous 
phase after deformation. Figure \ref{energycomparison}(b) shows the elastic strain energy stored in the crystalline (blue) and amorphous (red) phases, 
respectively. Before $\tilde{t}=5$, the elastic energy stored in the crystalline phase is much smaller that that in the amorphous phase, indicating that 
the crystalline phase is more stable. Under high stress, the crystalline phase becomes elastically unstable, leading to a rapid decrease in the elastic 
strain energy of the amorphous phases and an increase in the crystalline phase ($5<\tilde{t}<7$). After $\tilde{t}=7$, the amorphous phase becomes more stable and remain dominant in this 
region, which is aligned with Figure \ref{shearresults1}(c) and (d).

The simulation results demonstrate that amorphous shear bands form as a result of stress-driven propagation of the amorphous phase along high-stress 
trajectories. The presence of initial amorphous defects facilitates the nucleation and growth of shear bands, highlighting the critical role of microstructural 
imperfections in driving amorphization. Additionally, the energy analysis reveals that amorphization serves as an effective mechanism for dissipating 
energy in highly deformed materials, especially when plastic deformation is inhibited due to strain strengthening in the crystalline phase.

\subsection{Case 2: Grain size effect}

\begin{figure}[htbp]
    \centering
    \includegraphics[width=0.6\textwidth]{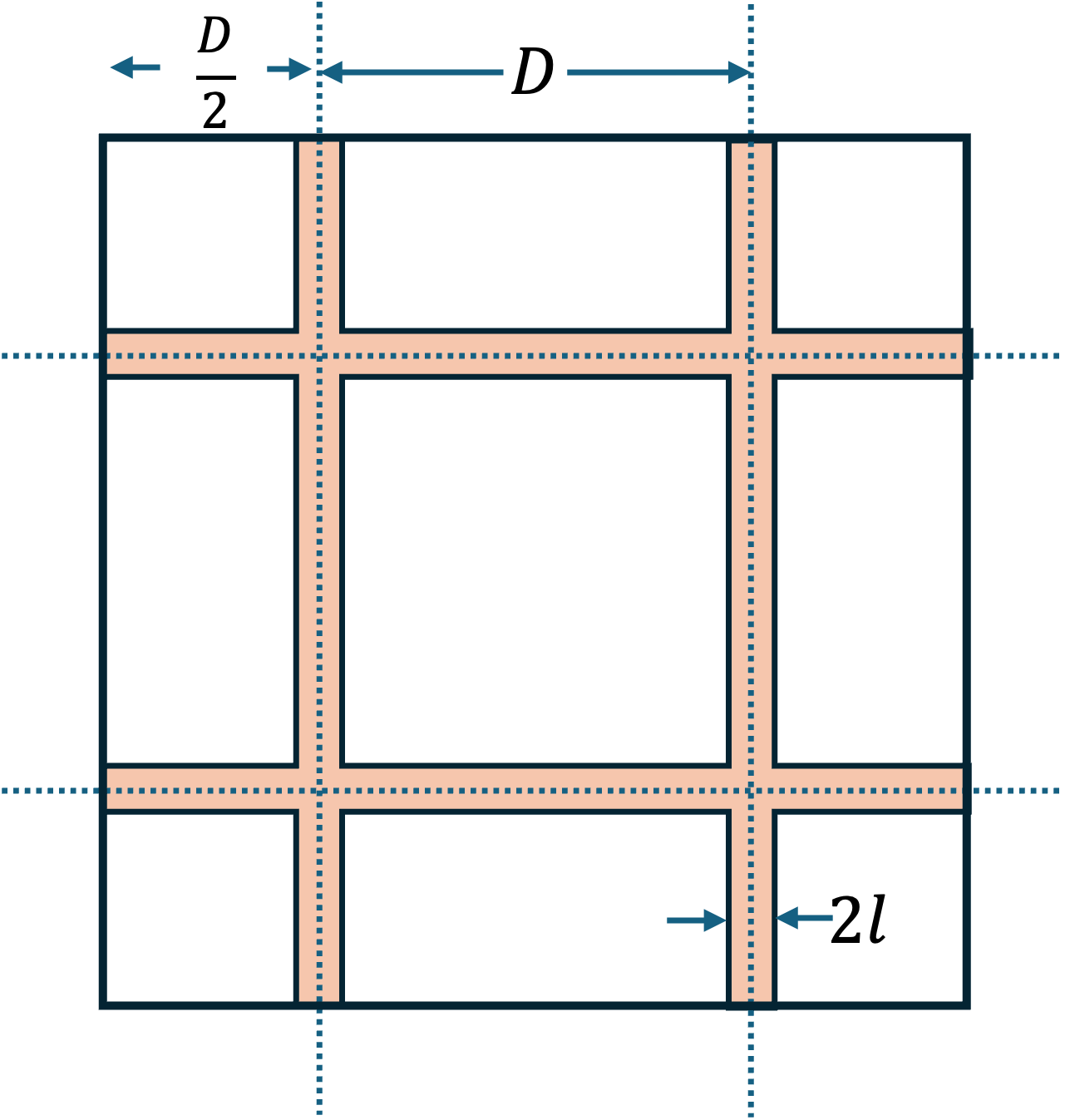}
    \caption{Multi-grain simulation setting. $D$ is the dimensionless size of grains and $2l$ is the dimensionless width
    of amorphous grain boundary, which is fixed.}
    \label{Grainsize_setting}
\end{figure}

In order to investigate the effect of grain size on amorphization, we simulate a multi-grain system inspired by \citet{xuGeneralizationHallPetchInverse2023}. 
The simulation domain consists of a polycrystalline structure with grain size $D$ and a fixed amorphous grain boundary width of $2l$ 
(Figure \ref*{Grainsize_setting}). To isolate grain size as the sole variable, crystallographic anisotropy is neglected, and amorphous 
grain boundaries are assumed to be identical. All material parameters are the same as those used in the previous case. 

\begin{figure}
    \centering
    \includegraphics[width=\textwidth]{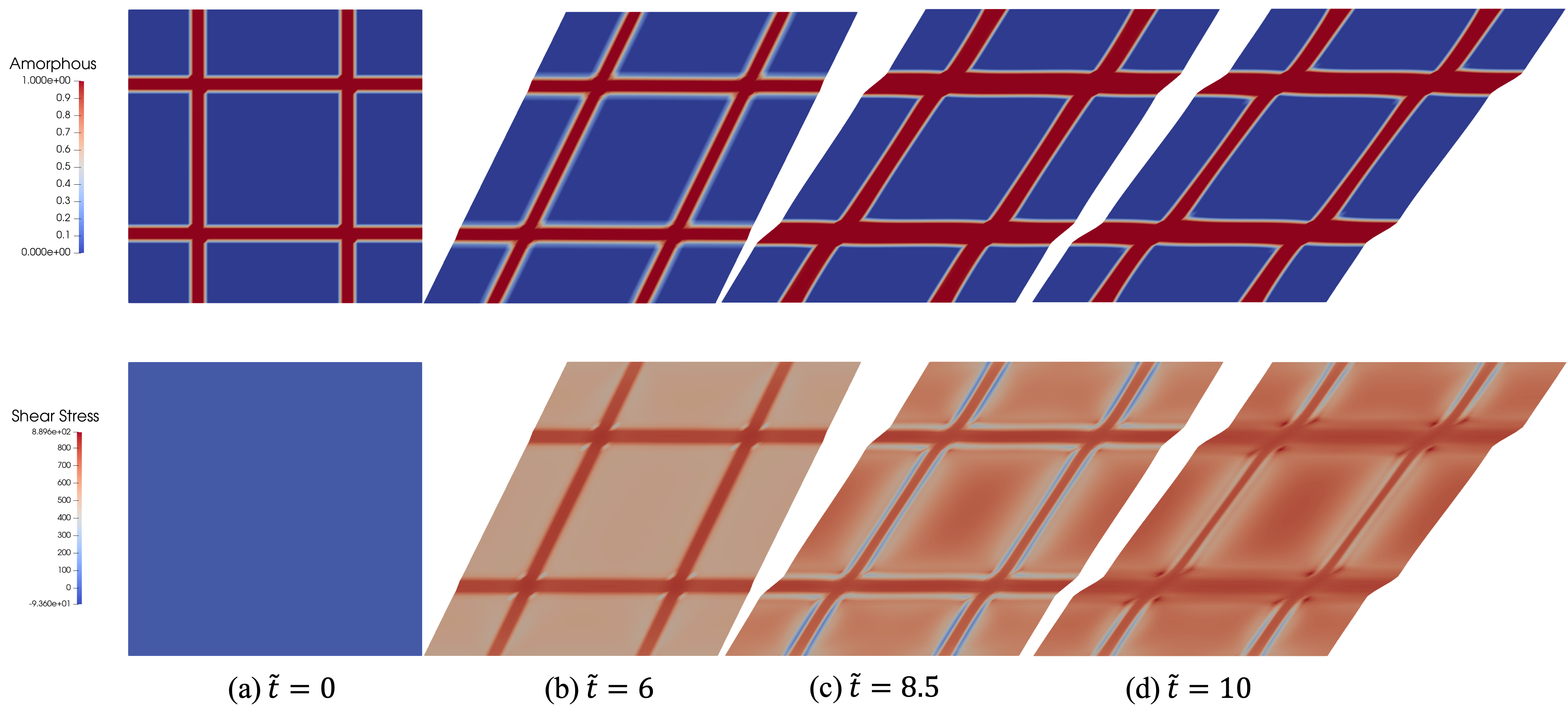}
    \caption{Evolution of the polycrystalline structure for $D=30$. The first row shows the propagation of amorphous grain boundaries,
    and the second row presents the evolution of the shear stress $\sigma_{xy}$.}
    \label{GrainSizeEvolution}
\end{figure}

We perform six simulations with grain sizes ranging from $D=20$ to $D=70$ and $l=1$ is chosen. Figure \ref{GrainSizeEvolution} shows the 
evolution of the polycrystalline structure for $D=30$. The first row illustrates the propagation of amorphous grain boundaries, while the 
second row shows the evolution of the shear stress $\sigma{xy}$. As deformation progresses, amorphous phases nucleate and propagate along 
grain boundaries, leading to the formation of shear bands.

\begin{figure}
    \centering
    \subfigure[Volume fraction]{\includegraphics[width=0.48\linewidth]{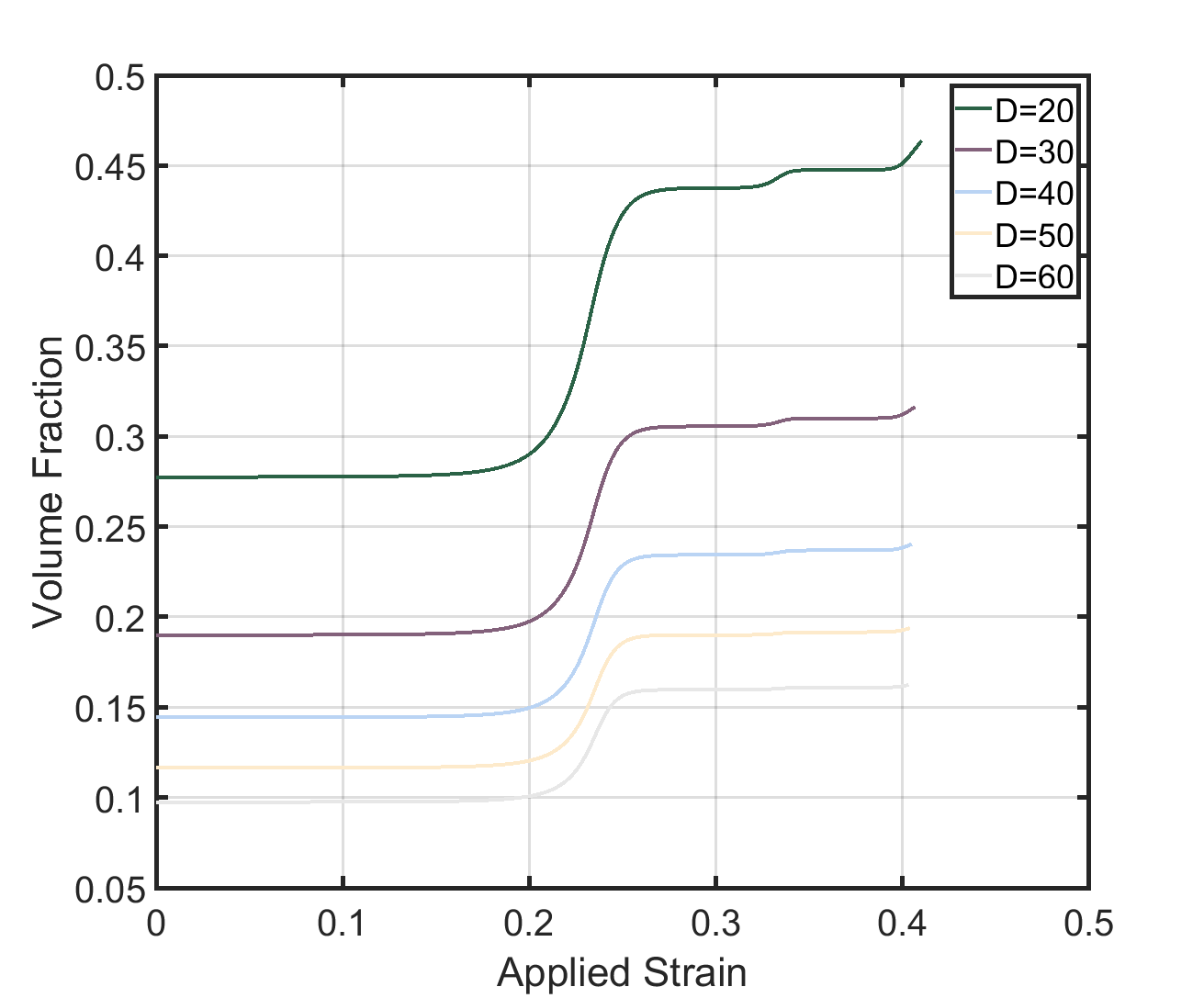}}
    \subfigure[Increase of Volume fraction]{\includegraphics[width=0.48\linewidth]{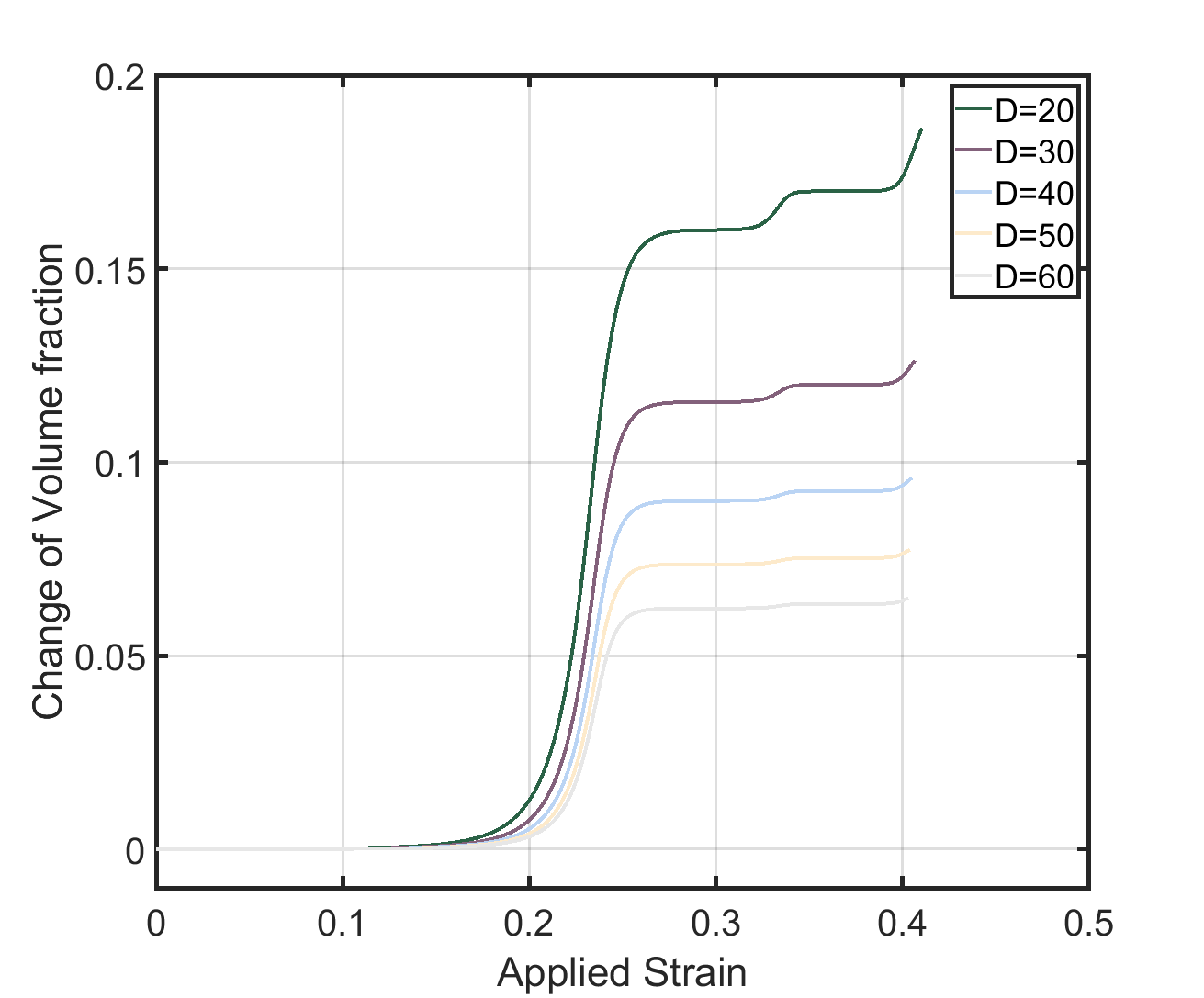}}
    \caption{Grain size effect on amorphization. (a) Volume fraction of amorphization ($V_g$) under large strains. (b) 
    Increase of the volume fraction of amorphous phases ($V_g-V_g^0$), where $V_g^0$ is the initial volume fraction.}
    \label{GrainSizeEffect}
\end{figure}

Figure \ref{GrainSizeEffect} summarizes the grain size effect on amorphization. In Figure \ref{GrainSizeEffect}(a), the volume fraction of
amorphous phases ($V_g$) increases with applied strain. Smaller grains exhibit a higher initial volume fraction of amorphous phases and a more 
significant increase in $V_g$ during deformation (Figure \ref{GrainSizeEffect}(b), where the increase in volume fraction of amorphous phases 
is presented). The results indicate that smaller grains facilitate the development of amorphous phases, consistent with experimental findings 
\cite{fanGrainSizeEffects2018, huaShearinducedAmorphizationNanocrystalline2022}. 

The strain-volume fraction curve exhibits horizontal strain jumps connected by nearly vertical volume fraction steps, as shown in Figure 
\ref{GrainSizeEffect}(a). This avalanche-like behavior, observed numerically for the first time, suggests that amorphization occurs in discrete bursts. 
During an avalanche, new amorphous phases form rapidly, dissipating energy and reducing the stress for further amorphization.  This behavior is 
analogous to avalanche dynamics commonly observed in plastic deformation, indicating that amorphization may be a critical-like process governed 
by similar dynamics.

The change in volume fraction of amorphous phases demonstrate that the threshold strain for amorphization increases slightly with grain size. 
In Figure \ref{GrainSizeEffect}(b), the curves for different grain sizes are close at the beginning, indicating that the stress condition 
dominates the nucleation of amorphous phases. For larger grains, the reduced number of nucleation sites slightly increases the threshold strain, 
but the change is minimal within the simulated grain size range.

\begin{figure}
    \centering
    \subfigure[Stress v.s. Strain]{\includegraphics[width=0.48\linewidth]{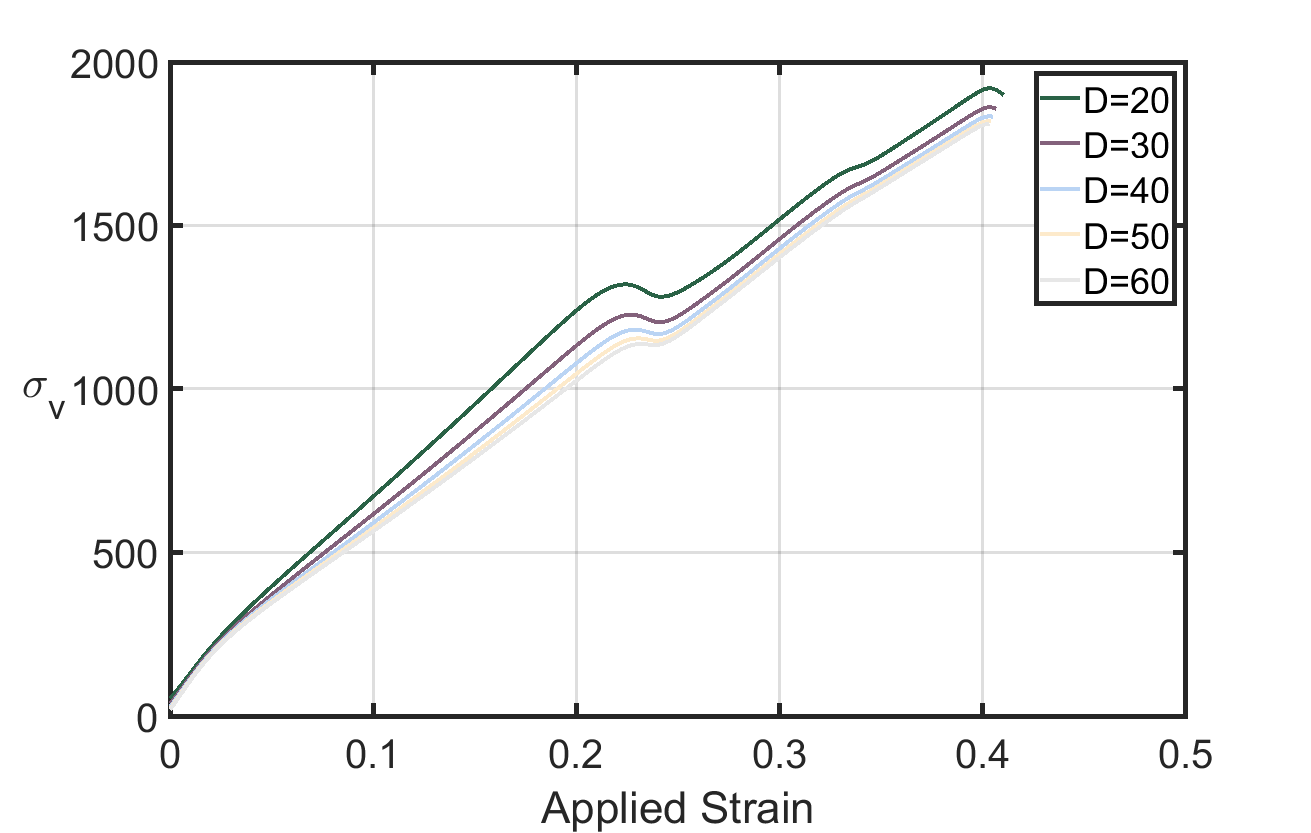}}
    \subfigure[Yield Stress v.s $D^{-\frac{1}{2}}$]{\includegraphics[width=0.48\linewidth]{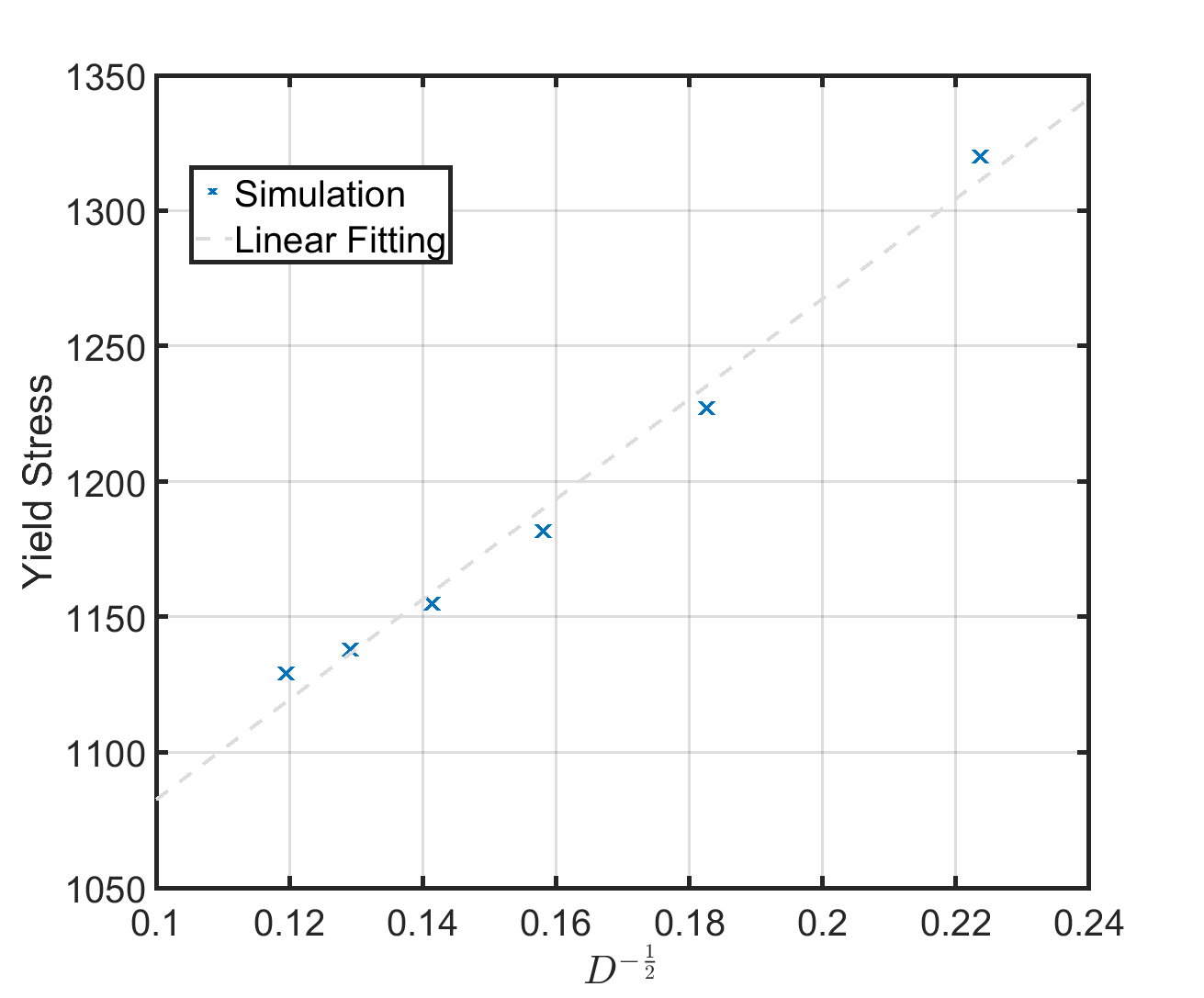}}
    \caption{Hall-Petch effect. (a) Stress-strain curves for different grain sizes. (b) Yield stress as a linear function of 
    $D^{-\frac{1}{2}}$. The blue line represents the linear fitting.}
    \label{HallPetcheffect}
\end{figure}

Figure \ref{HallPetcheffect} shows the stress-strain curves and yield stress as a linear function of $D^{-1/2}$. The yield stress increases 
with decreasing grain size, demonstrating the famous Hall-Petch effect. This observation validates the proposed phase-field model, as it captures 
the Hall-Petch behavior even in the absence of dislocations, highlighting the importance of continuum approaches for studying amorphization.

\subsection{Case 3: Surface amorphization}

To investigate surface amorphization under compression, we perform 3D simulations using MOOSE \cite{lindsay2022moose}. The simulation domain is 
a cubic cell with a side length of $L=170\, \text{nm}$. A random initial value between 0 and 0.1 is assigned to the phase variable $\eta$
to represent the initial amorphous defects. 

Compression is applied along the $Z$-axis to the nanocrystalline NiTi shape-memory alloy, The simulation parameters are the same as those used in
the previous cases. 

\begin{figure}[htbp]
    \centering
    \includegraphics[width=0.9\textwidth]{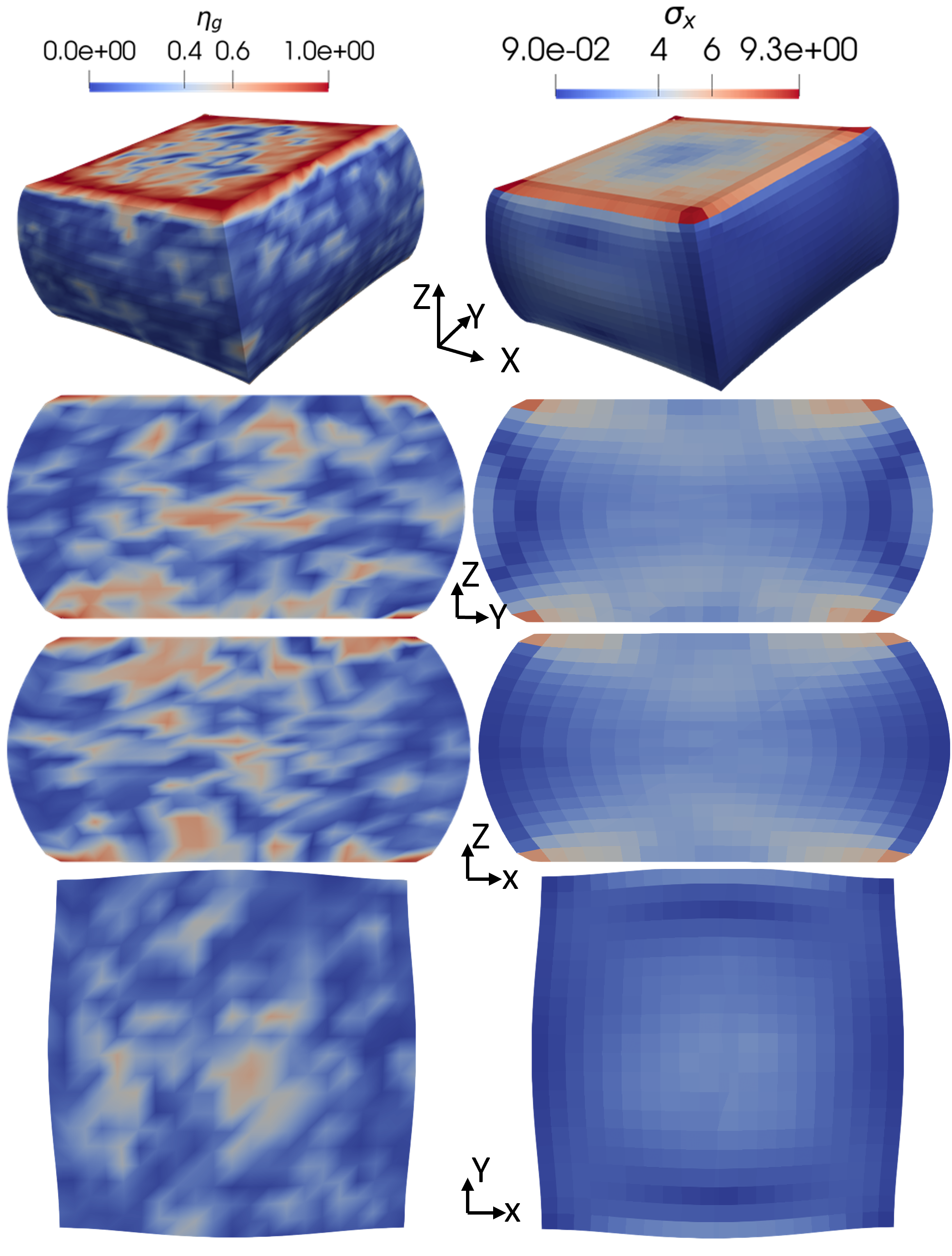}
    \caption{Results of compression in three dimensions. The first row illustrates the distributions of the order 
    parameter $\eta$ (left), and the stress magnitude $\sigma_x$ (right). The following rows show the cross-sectional views 
    along the Y-Z, X-Z, and X-Y planes.}
    \label{clipscompression}
\end{figure}

Figure \ref*{clipscompression} presents the results of the 3D compression simulation. The first row in Figure \ref*{clipscompression} 
illustrates an overview of the amorphous phase distribution (left) and the stress magnitude $\sigma_x$ (right). The subsequent rows show cross-sectional 
views along the Y-Z, X-Z, and X-Y planes. The phase variable $\eta$ highlights the amorphous regions,, while $\sigma_x=\sqrt{\sigma_{xx}^2+\
sigma_{xy}^2+\sigma_{xz}^2}$ represents the stress distribution. The results indicate that amorphous phases nucleate primarily on surfaces and 
in interior regions under high stress. 

\begin{figure}
    \centering
    \subfigure[$\varepsilon_c=0.228$]{\includegraphics[width=0.48\linewidth]{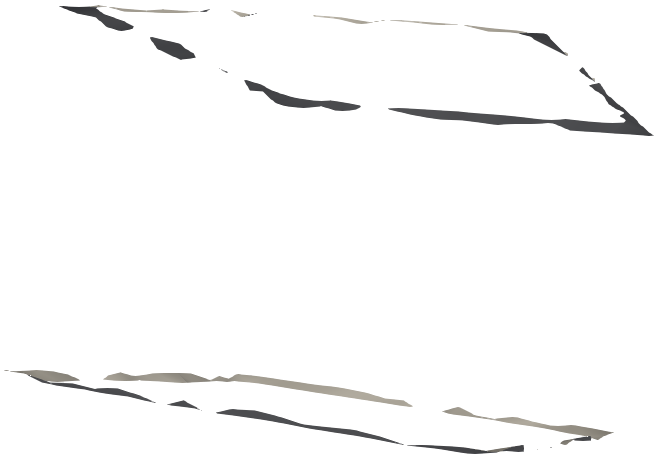}}
    \subfigure[$\varepsilon_c=0.302$]{\includegraphics[width=0.48\linewidth]{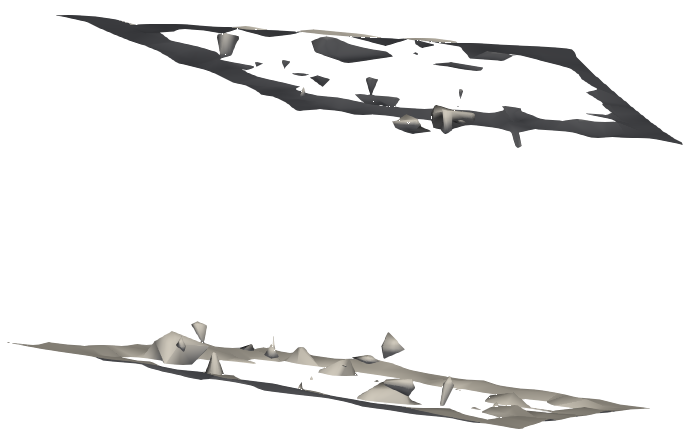}}\\
    \subfigure[$\varepsilon_c=0.332$]{\includegraphics[width=0.48\linewidth]{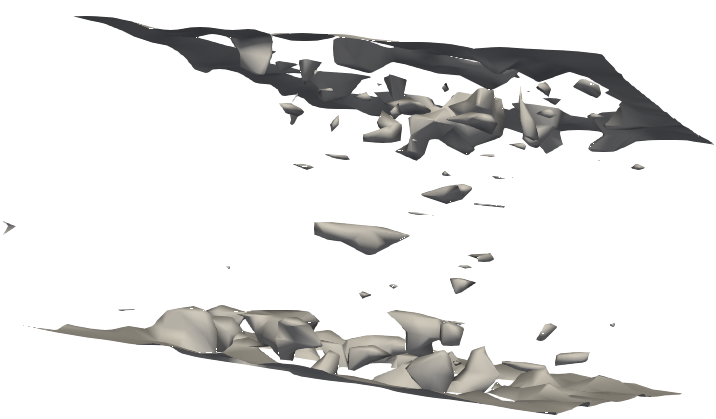}}
    \subfigure[$\varepsilon_c=0.365$]{\includegraphics[width=0.48\linewidth]{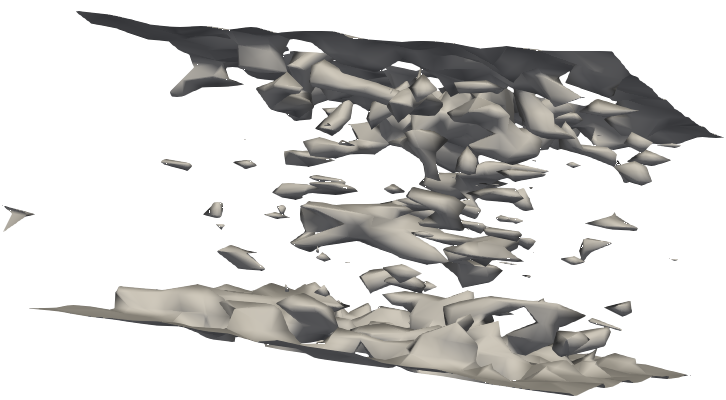}\label{contourd}}
    \caption{Isosurfaces of $\eta=0.5$ under compression. (a) and (b) show amorphous phases forming on surfaces at lower strains. 
    (c) and (d) illustrate the expansion of amorphous regions into diagonal and highly distorted areas at higher strains.}
    \label{contourscompression}
\end{figure}

Figure \ref*{contourscompression} shows the isosurfaces of $\eta=0.5$, which is considered as a threshold for amorphization. At lower compression
strain ($\varepsilon_c\leq 0.302$), amorphous phases form predominantly on the surfaces (Figure \ref{contourscompression}(a) and (b)). As the 
strain increases, the amorphous regions expand and align with the diagonal regions and surfaces of the compressed cell (Figure \ref{contourscompression}(c) 
and (d)).These regions correspond to areas of high distortion and are consistent with experimental observations 
\cite{zhaoAmorphizationNanocrystallizationSilicon2016, guoExtremelyHardAmorphouscrystalline2018, zhaoAmorphizationExtremeDeformation2021}. 

The results highlight the role of stress in driving amorphization, complementing the findings from the previous cases. While Case 1 demonstrated 
the formation of amorphous shear bands and Case 2 revealed the grain size dependence of amorphization, this case emphasizes the importance of surface 
and geometric effects. Together, these insights underscore the versatility of the proposed model in capturing the key mechanisms of stress-induced 
amorphization and its potential for studying amorphization as a deformation mechanism in nanocrystalline materials at the mesoscale.


\section{Conclusions}

This study introduces a novel phase-field model to investigate amorphization under severe plastic deformation. By integrating 
elastoplastic theory with phase field approaches, the model incorporates a new deviatoric stress-dependent transformation strain tensor, 
which captures the stress-driven nature of amorphization. This framework couples transformation strain energy with plastic work, enabling the 
simulation of amorphous phase nucleation and propagation under high stress. The proposed model provides a robust and thermodynamically consistent 
tool for studying amorphization as a deformation mechanism in nanocrystalline materials.

The numerical simulations validate the model and provide new insights into stress-induced amorphization. The formation and propagation of amorphous 
shear bands, driven by elastic instability in the crystalline phase, align with experimental observations and highlight the critical role of stress 
in driving amorphization. The grain size effect is also captured, demonstrating that smaller grains facilitate amorphization, while larger grains 
increase the critical strain required for amorphous phase formation. This behavior is consistent with the Hall-Petch effect, further validating 
ability of the model to reproduce fundamental material behaviors.

A particularly novel discovery is the avalanche-like dynamics of stress-induced amorphization, characterized by discrete bursts of amorphous phase 
formation. This critical-like behavior underscores the interplay between amorphization and plasticity, offering a new perspective on the mechanisms 
governing deformation in nanocrystalline materials. Additionally, 3D simulations reveal that amorphous phases nucleate at surfaces 
and stress concentrators, such as diagonal regions in compressed cells. These findings are consistent with in situ experimental studies and emphasize 
the importance of surface and geometric effects in amorphization.

Together, these results demonstrate the versatility of the proposed model in capturing the key mechanisms of stress-induced amorphization. By reproducing 
experimentally observed phenomena, such as the Hall-Petch effect and surface amorphization, the model provides a quantitative framework for understanding 
the interplay between stress, defects, and phase transformation. These insights have significant implications for material design, particularly for 
developing materials with enhanced resistance to deformation-induced amorphization under extreme mechanical conditions.

Looking ahead, the model can be extended to include additional deformation mechanisms and defects, such as dislocations and grain boundary evolution, which 
play a critical role in amorphization. Incorporating thermal effects and temperature-dependent material properties will further enhance applicability of the 
proposed model. Large-scale simulations and experimental validation will also be pursued to bridge the gap between mesoscale modeling and macroscale material 
behavior. This work lays the groundwork for advancing predictive models of amorphization and provides a powerful tool for studying the underlying mechanisms 
of deformation-induced amorphization in crystalline materials.

\section*{Acknowledgement}
This work was supported by the Hong Kong Research Grants Council Collaborative Research Fund C6016-20G and the Project of Hetao 
Shenzhen-HKUST Innovation Cooperation Zone HZQB-KCZYB-2020083.

\begin{appendices}
\section*{Appendix: Symbols }
\setlength{\tabcolsep}{4pt}

\begin{longtable}{p{3cm}p{12cm}}
    \caption{List of symbols in finite strain framework}\\ 
        \hline
        Symbol                  &    Meaning                                              \\
        \hline
        \endfirsthead

        \hline
        \endfoot

        \hline
        Symbol                  &    Meaning                                              \\
        \hline
        \endhead

        \hline 
        \endfoot

        \hline
        Symbol                  &    Meaning                                              \\
        \hline
        \endhead

        \hline
        \endlastfoot

        $\eta$              &  Order parameter              \\
        $V_g$               &  Volume fraction of the amorphous phase \\
        $\rho$              &  Mass density of the system \\
        $\rho_c$/$\rho_g$   &  Mass density of the crystalline/amorphous phase \\
        $J_g$             &  Mass density ratio between the crystalline and amorphous phase \\
        $\Omega_0$/$\Omega$             &  Reference/deformed configuration                              \\
        $\bm{x}$                &  Material point in $\Omega_0$                         \\
        $\bm{X}$                &  Image point in $\Omega$                            \\
        $\bm{F}$                &  Total deformation gradient                         \\
        $\bm{F}_c$/$\bm{F}_g$   &  Deformation gradient in the crystalline/amorphous phase               \\
        $\bm{F}_e^c$/$\bm{F}_e^g$   &  Elastic deformation gradient in the crystalline/amorphous phase \\
        $\bm{F}_p^c$/$\bm{F}_p^g$   &  Plastic deformation gradient in the crystalline/amorphous phase \\
        $\bm{F}_t$              &  Transformation gradient                           \\
        $\bm{L}_p^c$/$\bm{L}_p^g$   &  Plastic velocity gradient  in the crystalline/amorphous phase                \\
        $N_s$                   &  Number of slip systems in crystal                  \\
        $\alpha,\beta$          &  Indices of slip systems                                \\
        $\dot{\gamma}^{\alpha}$ &  Shear rate on the slip system $\alpha$                 \\
        $\bm{m}^{\alpha}$       &  Slip direction on the slip system $\alpha$             \\
        $\bm{n}^{\alpha}$       &  Slip normal of the slip system $\alpha$                \\
        $\tau^{\alpha}$         &  Resolved stress on the slip system $\alpha$        \\
        $\dot{\gamma}_0$        &  Reference shear rate                 \\
        $m$                     &  Strain rate sensitivity                           \\
        $\tau_c^{\alpha}$       &  Slip resistance on the slip system $\alpha$        \\
        $h_{\alpha\beta}$       &  Hardening matrix                                   \\
        $\hat{\tau}^g$            &  Deviatoric second Kirchhoff stress tensor                 \\
        $\bm{N}$                &  Visco-plastic flow vector                          \\
        $\dot{\gamma}_g$        &  Visco-plastic multiplier                           \\
        $\bm{R}_e$              &  Elastic rotation tensor                            \\
        $A$                     &  A material parameter        \\
        $\tau_{eq}^g$           &  Kirchhoff equivalent stress tensor                        \\
        $\tau^*$                &  Reference stress for amorphous phase               \\
        $\dot{\gamma}_t$        &  Reference rate of transformation strain           \\
        $\psi$                  &  Total energy functional density                           \\
        $\psi^{ch}$      &  Local phase separation energy density                      \\
        $\psi^{\nabla}$        &  Gradient energy density                                  \\
        $\psi_e^c$/$\psi_e^p$            &  Elastic strain energy density in the crystalline/amorphous phase                           \\
        $A,B,C$                &  Parameters determine local phase separation energy   \\
        $\kappa$               &  Energy barrier between the crystalline and amorphous phase    \\
        $\beta$       &  Coefficients related to the interfacial energy         \\
        $\bm{E}_e^c$/$\bm{E}_e^g$       &  Elastic strain tensor in the crystalline/amorphous phase                            \\
        $\mathbb{C}_c$/$\mathbb{C}_g$   &  Elastic coefficients in the crystalline/amorphous phase            \\
        $\bm{P}^c$/$\bm{P}^g$       &  First Piola-Kirchhoff stress tensor in the crystalline/amorphous phase \\
        $\bm{\sigma}^c$/$\bm{\sigma}^g$       &  Cauchy stress tensor in the crystalline/amorphous phase \\
        $\hat{\bm{\sigma}}^c$/$\hat{\bm{\sigma}}^g$       &  Second Kirchhoff stress tensor in the crystalline/amorphous phase \\
        $\bm{P}$                &  First Piola-Kirchhoff stress tensor                \\
        $\bm{X}_p^c$/$\bm{X}_p^g$      &  Work-conjugate force for plastic strain in the crystalline/amorphous phase \\
        $\bm{X}_t$/$\bm{X}_{\eta}$     &  Work-conjugate force for transformation strain and phase field variable \\
        $M$               &  Mobilities for amorphization            \\
\end{longtable}

\end{appendices}

\bibliography{CitedReference}

\end{document}